# Tuning chemical short-range order for stainless behavior at reduced chromium concentrations in multi-principal element alloys

W.H. Blades[1], B.W.Y. Redemann[2,3,4], N. Smith[5], D. Sur[6], M.S. Barbieri[6], Y. Xie[1], S. Lech[3], E. Anber[3], M.L. Taheri[3], C. Wolverton[5], T.M. McQueen[2,3,4], J.R. Scully[6], K. Sieradzki[1*]

[1]Ira A. Fulton School of Engineering; Arizona State University; Tempe, AZ, 85287, USA.

[2]Department of Chemistry, The Johns Hopkins University; Baltimore, MD, 21218, USA.

[3]Department of Material Science and Engineering, The Johns Hopkins University; Baltimore, MD, 21218, USA.

[4]William H. Miller III Department of Physics and Astronomy, The Johns Hopkins University; Baltimore, MD, 21218, USA.

[5]Department of Materials Science and Engineering, Northwestern University; Evanston, IL, 60208, USA

[6]Department of Material Science and Engineering, University of Virginia; Charlottesville, VA, 22904, USA.\

## Abstract

Single-phase multi-principal element alloys (MPEAs) hold promise for improved mechanical properties as a result of multiple operative deformation modes. However, the use of many of these alloys in structural applications is limited as a consequence of their poor aqueous corrosion resistance. Here we introduce a new approach for significantly improving the passivation behavior of alloys by tuning the chemical short-range order (CSRO) parameter. We show that the addition of only 0.03 to 0.06 mole fraction of Al to a $(FeCoNi)_{0.9}Cr_{0.1}$ alloy changed both the magnitude and sign of the Cr-Cr CSRO parameter resulting in passivation behavior similar to 304L stainless steel containing twice the amount of Cr. Our analysis is based on comparing electrochemical measures of the kinetics of passive film formation with CSRO characterizations using time-of-flight neutron scattering, cluster expansion methods, density functional theory and Monte Carlo techniques. Our findings are interpreted within the framework of a recently proposed percolation theory of passivation that examines how selective dissolution of the non-passivating alloy components and CSRO results in excellent passive films at reduced levels of the passivating component.

*Correspnding author: karl.sieradzki@asu.edu

# 1. Introduction

Multi-principal elements alloys (MPEAs) that are variously termed high entropy or compositionally complex alloys, hold promise for improved mechanical and corrosion properties. There is now considerable experimental evidence that some single-phase MPEAs show enhanced yield strength, ductility and fracture toughness[1-6]. The situation regarding MPEA corrosion behaviors is far less clear[7-12]. Generally single-phase alloys show better corrosion behavior than a multi-phase counterpart at virtually the same composition since a secondary phase can result in enhanced pitting behavior and micro-galvanic corrosion. Equimolar MPEAs containing six components or fewer, in which Cr serves as one of the components, should show good aqueous passivation behavior unless one of the other components is particularly electrochemically active. Such an active component will not be able to participate in the formation of a thin (~ 2 nm) mixed oxide/hydroxide passive film and will serve to destabilize the film by electrochemical dissolution[8]. In a six-component equimolar alloy the Cr mole fraction is in the range of 0.16-0.17 which has been shown to be sufficient in the case of Fe-Cr and Ni-Cr binary alloys for excellent passive film behavior[13]. A key criterion for this is the ability of the film to self-heal with little accompanying metal dissolution[13]. If there is an electrochemically active component in the equimolar alloy, metal dissolution accompanying passive film formation will be large. The single-phase equimolar FCC Cantor alloy (CoCrFeMnNi) is an example of a MPEA that displays excellent mechanical properties [1,3-5], but the passive film behavior of the alloy is poor since the Mn component is too electrochemically active to contribute to the stability of the passive film[7,8].

Generally single-phase metallic alloys containing a high enough mole fraction of a component that forms a protective passive film in its elemental state, such as Cr, will result in enhanced alloy corrosion properties. A recent theory of passivation based on percolation concepts focuses on the role of site percolation of the passivating component(s) across the surface of single-phase alloys[13]. Chemical short-range order (CSRO) of the "clustering" type was found to significantly reduce percolation thresholds and so promote the formation of a protective passive film at a reduced mole fraction of a passivating component such as Cr. The theory predicts the number of atomic layers, $h$, that have to undergo selective dissolution of the non-passivating component(s) in order for Cr to percolate across the corrosion roughened surface. Values of $h$ in the range of 5-20 atomic layers are indicative of excellent passivation behavior.

The typical approach to tune CSRO in an alloy involves changes in alloy chemistry, combined with thermal processing at temperatures above that which would result in compound formation or phase separation. Since these changes in alloy composition and processing temperatures are coupled, an important question to ask is, can the addition of a component to a non-equiatomic MPEA simultaneously maintain a single phase, act in a manner to promote CSRO, and allow for passivation at reduced levels of Cr? We investigated this by adding Al to $(FeCoNi)_{1-x}Cr_x$ alloys (x represents the mole fraction). We performed a series of electrochemical tests, X-ray photoelectron spectroscopy (XPS) to characterize the composition of the passive films and time-of-flight neutron scattering to determine the CSRO. Computational methods were employed based on cluster expansion, density functional theory and Monte Carlo methods aimed at determining whether the addition of Al can be used to tune the CSRO. These alloys were chosen since various aspects of the equimolar ternary NiCoCr and quaternary FeCoNiCr have been previously investigated including the CSRO and mechanical properties[14-18].

It is important to highlight the complicated role of Al in the passivation of single-phase metallic alloys. Unlike Cr, the passivation behavior of elemental Al in $H_2SO_4$ and other acids is poor [19,20]. Nevertheless, in single-phase binary alloys such as Fe-Al, an Al mole fraction of 0.20 or greater significantly improves passivation behavior [19,21]. In a random body-centered cubic (BCC) lattice, at this composition, the average Al cluster size including 1st and 2nd nearest neighbors is a trimer that will form an oxide cluster as a result of the oxyphilic nature of Al [13,22]. These clusters serve as nucleation centers for the incorporation of the surrounding iron atoms into a mixed oxide, thus bypassing the normal salt-film passivation mechanism of iron in sulfuric acid [23].

## 2. Methods

### 2.1 Alloy preparation

(FeCoNi)$_{1-x}$Cr$_x$ and (FeCoNi)$_{1-x-y}$Cr$_x$Al$_y$ (x and y represent mole fractions) alloys were prepared by induction melting using a 50 kW power source at 150 kHz in a water-cooled copper crucible under a He atmosphere using high purity raw metals (> 99.95%, Neyco). The (FeCoNi) compositions were nominally equimolar. To determine the separate effects of Cr and Al on the passivation processes, two sets of "control" (FeCoNi)$_{1-x}$Cr$_x$ and (FeCoNi)$_{1-y}$Al$_y$ alloys were fabricated by arc melting using pure elements: 99.95% Fe, 99.95% Co, 99.995% Ni and 99.99% Al (Kurt J. Lesker Company). All as-cast samples were then encapsulated in quartz tubes with a forming gas (95% Ar + 5% $H_2$ atmosphere) and Ta foil getter. Each set of alloys was homogenized at 1,000 °C for 48 h by sealing each ingot in an evacuated fused silica tube (14 mm ID, 16 mm ID), heating rapidly to 1000˚ C (by placing in a preheated furnace), followed by ice water quenching after 48 hr. After quenching, the bulk ingots were cut using a water jet cutter (ProtoMAX Personal Abrasive Waterjet with 85HPX granite abrasive) and hand-polished using 1200 grit SiC polishing paper to remove the cut surface along the ingots. The alloy compositions and homogeneity were confirmed with energy dispersive spectroscopy using a Helios UC G4 SEM. Before electrochemical experimentation, each alloy surface was polished with a diamond suspension to a 1 μm finish.

The 304L stainless steel is a commercial grade (McMaster-Carr) with the composition Fe$_{0.692}$Ni$_{0.074}$Cr$_{0.20}$Mn$_{0.02}$Si$_{0.014}$ (subscripts represent mole fraction).

### 2.2 X-ray diffraction

X-ray diffraction (XRD) was used to identify the phases present in a set of FeCoNi-Cr-Al alloys with varying compositions. Diffraction data was collected on a Malvern PANalytical Empyrean diffractometer. All scans were performed using Cu Kα radiation at a voltage of 45kV and a current of 40 mA. Scans for most alloy samples were performed using an area of 36 mm$^2$ at a rate of 3º min$^{-1}$ over a 2θ scanning range of 20 -120º. A line scan was used for these samples. Scans for the Al control alloys, (FeCoNi)$_{0.97-0.87}$Al$_{0.03-0.13}$, were performed using an area of 26 mm$^2$ at a rate of 2.15º min$^{-1}$ over a 2θ scanning range of 20 -120º. A smaller area was used since the sample set had a slightly smaller cross-sectional area. An area scan was used for these samples to improve the signal to noise ratio, and as a result reduced the effects that texture may have had on peak intensity. Due to the presence of Fe and Co, which fluoresce under Cu radiation, the proportional height detection levels of the detector were adjusted to have a range of 50% - 75% to improve the signal-to-noise ratio. Scans shown in Figure S1were stripped of Kα-2 peaks.

## 2.3 *Scanning electron microscopy/Energy dispersive spectroscopy*

All MPEA compositions were measured with a Helios UC G4 scanning electron microscope using energy dispersive spectroscopy. An Everhart-Thornley detector was used in both secondary electron and backscatter electron modes with a beam current,.of 3.2 nA and a beam voltage of 15 kV or 18kV. The EDS spectra were identified and formatted using the Oxford Instruments AZtec software package.

The size of the $2^{nd}$ phase $\beta$-NiAl precipitate particles in the $(FeCoNi)Cr_{0.10}Al_{0.09}$ alloy and their total areal coverage was characterized in a backscattered mode. SEM micrographs were taken using an Ion Conversion and Electron (ICE) detector at a 18 kV accelerating voltage, beam current of 3.2 nA and a dwell time of 10 μs. These images were then post-processed using a custom MATLAB code by setting a minimum greyscale threshold value and removing all pixels below this threshold (i.e. the image background), Small image artifacts below 8 pixels in size were automatically removed by the MATLAB code and larger defects (e.g. polishing marks) were manually removed post segmentation. The area of those pixels representing the B2 phase was quantitatively determined and their distribution plotted.

## 2.4 *TEM/STEM and electron diffraction*

The as-homogenized ingots were sliced into a 300 μm thick samples by electrical discharge machining followed by grinding to a final thickness of 80 μm using standard metallographic procedures. Then, 3 mm discs were punched for the final electropolishing. It was performed using a twin-jet Struers Tenupol 5 electropolishing system using the electrolyte of 5% perchloric acid in methanol at -40 ℃ until a perforation was formed in the disc.

Transmission and scanning transmission electron microscopy (TEM/STEM) was used to identify the phases in a set of as-homogenized $(FeCoNi)_{0.90-y}Cr_{0.10}Al_y$ alloys. For the TEM/STEM investigations, a JEOL F200 TEM equipped with a dual light-element energy-dispersive X-ray spectroscopy detector was used. Examinations were conducted at 200 kV using conventional bright field imaging and selected area electron diffraction (SAED). The EDS technique was used for chemical composition analysis in STEM (maps) and TEM (point) modes. Quantitative analysis was done using the standard-less ratio method corrected by the sample thickness and density. The SAED patterns were analyzed using the JEMS software.

## 2.5 *Electrochemical measurements*

Linear sweep voltammetry (LSV) and chronoamperometry were recorded in 0.1 M $H_2SO_4$ electrolyte using a Gamry Interface 1000E potentiostat. Prior to data acquisition, the solution was de-aerated by bubbling ultra-high purity (UHP) $N_2$ for 15 min. The cell contained a platinum mesh counter electrode and a mercury/mercurous-sulfate reference electrode (+640 mV vs. SHE). All potentials are reported with respect to the standard hydrogen electrode (SHE). After the alloy surface was exposed to de-aerated electrolyte, the air-formed oxide was cathodically reduced by applying a series of potentials; −0.76 V for 300 s, −1.26 V for 3 s; −0.76 V for 60 s and −0.36 V for 10 s. During this reduction protocol potential UHP $N_2$ was used to remove hydrogen gas bubbles from the sample surface. The LSV was performed by ramping the voltage from -200 mV to 1000 mV at 1 mV $s^{-1}$. During chronoamperometry, the voltage was stepped from -120 mV (-160 mV for 304L) to either 150 mV or 350 mV depending on the experiment. The voltage was held for 300 s, and the charge density determined by numerically integrating the current density as a function of time. The number of monolayers dissolved ($h$) during primary passivation was then calculated by following a procedure developed previously [13]. Here, we assume the FCC alloy surface is primarily comprised of (111) faces and the metal cations are oxidized to valences of $Fe^{2+}$, $Co^{2+}$, $Ni^{2+}$, $Cr^{3+}$ and $Al^{3+}$.

All AC electrochemical experiments were performed using a BioLogic SP-200 potentiostat. Electrochemical Impedance Spectroscopy (EIS) was performed after a 10 ks potentiostatic hold at 150 mV. Before each experiment, the native oxide film on the working-electrode surface was cathodically reduced following the same procedure as that used for the LSV and chronoamperometry. A 20 mV signal was used to record impedance from 100 kHz to 5 mHz, at a rate of 8 points per frequency decade. Single Frequency EIS (SF-EIS) was used to monitor the current and imaginary impedance for the entire duration of hold using a 30 Hz, 20 mV AC signal at an interval of 0.3 s. The |$Z$| value used to obtain the results in Figs. 2a and Fig. S8 was at a frequency of 5 mHz.

### 2.6 *XPS analysis*

A PHI VersaProbe-III$^{TM}$ XPS Analyzer was used to characterize the cationic species in the passive films The instrument was calibrated with a Au standard to the $4f_{7/2}$ core level at 84.00 eV binding energy. High resolution XPS spectra were collected using Al Kα X-rays (1,468.7 eV) at a 26 eV pass energy, 45° take off angle, 20 kV accelerating voltage, and a 100 x100 μm spot size. The scan energy and time per step were 0.05 eV and 50 ms, respectively. Relative atomic sensitivity factors as provided by PHI were used to normalize elemental contributions and obtain cationic compositions of the film. Due to the overlap of several Auger peaks for Fe and Co with their core level 2p3/2 ranges, the following core level regions were focused on in the analysis: Al 2p, Cr $2p_{3/2}$, Fe $2p_{1/2}$, Co $2p_{1/2}$, Ni $2p_{3/2}$ and O 1s. KOLXPD$^{TM}$ was used for spectral fitting. All peaks were charge shifted using C 1s = 284.80 eV and deconvoluted for metallic features, and Voigt functions for cationic features [24-26]. The fitting procedure followed Biesinger et al. for reference spectra for Fe, Co, Ni, and Cr and its species [24]. For Al, the NIST database for $Al_2O_3$ and $Al(OH)_3$ was used [26]. The small overlap of Cr 3s was resolved by fixing its amplitude and calculating it from the proportion of the Cr $2p_{3/2}$ core shell total signal. A set of initial parameters consisting of all the possible multiplets, following the reference spectra, was used and then a numerical fitting procedure was performed. The area under the peaks were obtained for each cationic species. The ($Cr^{+3}$ + $Al^{+3}$) mole fraction in Fig. 2d represents the sum of mole fractions of Cr oxides and hydroxides for $Cr^{3+}$ while $Al^{+3}$ represents $Al_2O_3$ and $Al(OH)_3$. The ($Cr^{+3}$ + $Al^{+3}$) mole fractions were determined by integrating the area under the Cr$2p_{3/2}$ and Al2p core shell binding energy peaks (Figs. S10-S13) and dividing by the instrument relative sensitivity factors (Table S1) of the peaks to obtain the fraction of $Cr^{+3}$, ($f_{Cr+3}$ and $Al^{+3}$, ($f$Al2p), provided in Table S2. Then for example, the mole fraction (mF) of $Cr^{+3}$ was determined according to: $mFCr^{+3} = f_{Cr+3}/\Sigma_M f_M$, where $\Sigma_M f_M$ is the sum of the fractions of all the metallic cation species. Table S3 shows binding energy and FWHM fitting parameters used for the core shells of Cr $2p_{3/2}$, Fe $2p_{1/2}$, Co $2p_{1/2}$, Ni $2p_{3/2}$, Al 2p, and O 1s peaks.

## 2.7 *Neutron Scattering*

Room temperature time of flight (ToF) powder neutron diffraction was performed at Oak Ridge National Laboratory, on the POWGEN diffractometer with Frame 1 ($\lambda_{center}$ = 0.8 Å). Initial data reduction was performed using Mantid version 6.5.0 [27] to obtain the $S(q)$, $F(q)$ and the pair distribution function (PDF, $G(r)/D(r)$) for analysis. Initial refinements of the diffraction data to an average FCC unit cell was performed using GSAS-II [28], and initial small box PDF refinements were performed using PDFgui [29]. Large box (containing 20 x 20 x 20 conventional FCC unit cells) reverse Monte Carlo (RMC) refinements of the PDF data were carried out using RMCProfile 6.7.9 [30-32]. The $G(r)$ and $S(q)$ were fit simultaneously, with the $G(r)$ more heavily weighted. The distribution of atom types within each 32,000 - atom box was fixed to enforce ratios corresponding to the nominal composition, and constraints were implemented on the maximum and minimum distances for the atomic pairs. Each RMC step allowed both atomic displacements as well as atom identity swaps. RMC refinements were performed ten times for each composition, with a randomly initialized starting configuration, and run for 24 hours on single 48-core nodes of the Rockfish supercomputer. Results were visualized and analyzed with Vesta, Crystal Maker version 10.8.1, and Python. The number of Cr-Cr pairs in the first nearest neighbor shell shown in Fig. 3c was determined from $\frac{\#\ of\ Cr-Cr\ neighbors}{\frac{4}{3}\pi(\frac{r+dr}{2})^3-\frac{4}{3}\pi(\frac{r-dr}{2})^3}$ , where $r$ =2.45 Å and $dr$ = 0.10 Å.

## 2.8 *Computational*

Density functional theory (DFT), Cluster Expansion (CE) and Monte Carlo (MC) techniques: Density Functional Theory calculations used to generate the training set were performed using VASP [33-36] utilizing the projector-augmented wave [36] method and GGA-PBE functionals [37]. All calculations were performed non-magnetically. The CE was then used to provide an effective Hamiltonian for canonical MC simulations to provide SRO parameters at finite temperature. The CE method has been used successfully to model materials for myriad applications such as metallic alloys, surfaces and nanoparticles [38,39]. The cluster expansion (CE) method is a generalized Ising model which relates configurations of atoms, σ, on a parent lattice to their associated enthalpies, $E(\sigma)$. This takes the form, $E(\sigma) = \sum_{\alpha} m_{\alpha} J_{\alpha} < T_{\alpha'}(\sigma) >_{\alpha}$ ,where the sum runs over all symmetrically distinct clusters α with $m\alpha$ multiplicities of these clusters [40-41]. $J_{\alpha}$ are the Effective Cluster Interactions (ECI) usually obtained during fitting to DFT calculations, and $\langle \Gamma\alpha'(\sigma)\rangle\alpha$ are the cluster functions selected to form an orthogonal basis. Here we use the Alloy Theoretic Automated Toolkit (ATAT) [40-42] to perform the CE and construct the Hamiltonian used in subsequent MC sampling to compute CSRO. The cross-validation score was less than 13 meV/atom and 63 structures were included in the training set with up to 8 atoms per unit cell. The ground state convex hull is shown in Fig. S9. MC is a stochastic method which relies on statistical sampling over many configurations to obtain averages of thermodynamic quantities. Hence, the utilization of the CE method allows for a more rapid calculation of the system energy than from first principles. Here, we employed the canonical MC implementation of ATAT [43], using 8,000 atom supercells at T = 3,000 K. The system was allowed to equilibrate for 1,000 MC steps, and thermodynamic averages were obtained over the successive 32,000 steps. T = 3000K is demonstrated to be sufficiently high temperature to obtain a single phase in the composition region studied here, as shown in Fig. S9.

## 3. Experimental Results

### 3.1 *Electrochemical and X-ray Photoelectron Spectroscopy (XPS) Characterizations of Passivation*

The crystal structure characterization, compositional analysis and the microstructure of the alloys fabricated are shown in Supplementary Figs. S1-S5. Except for (FeCoNi)$_{0.81}$Cr$_{0.10}$Al$_{0.09}$, all the alloys were single-phase solid solutions at ambient temperature with a FCC crystal structure. As shown in Fig.1, we characterized the passivation behavior of (FeCoNi)$_{1-x-y}$Cr$_x$Al$_y$ alloys and two sets of control samples, (FeCoNi)$_{1-x}$Cr$_x$ and (FeCoNi)$_{1-y}$Al$_y$ in 0.10 M H$_2$SO$_4$. Figure 1a shows the linear sweep voltammetry (LSV) behavior for all the sets of alloys and 304L stainless steel. The LSV of the Al control alloys display poor passivation behavior. The passivation wave is wide (~350 mV) indicating there is a large amount of dissolution required prior to the onset of passivation.

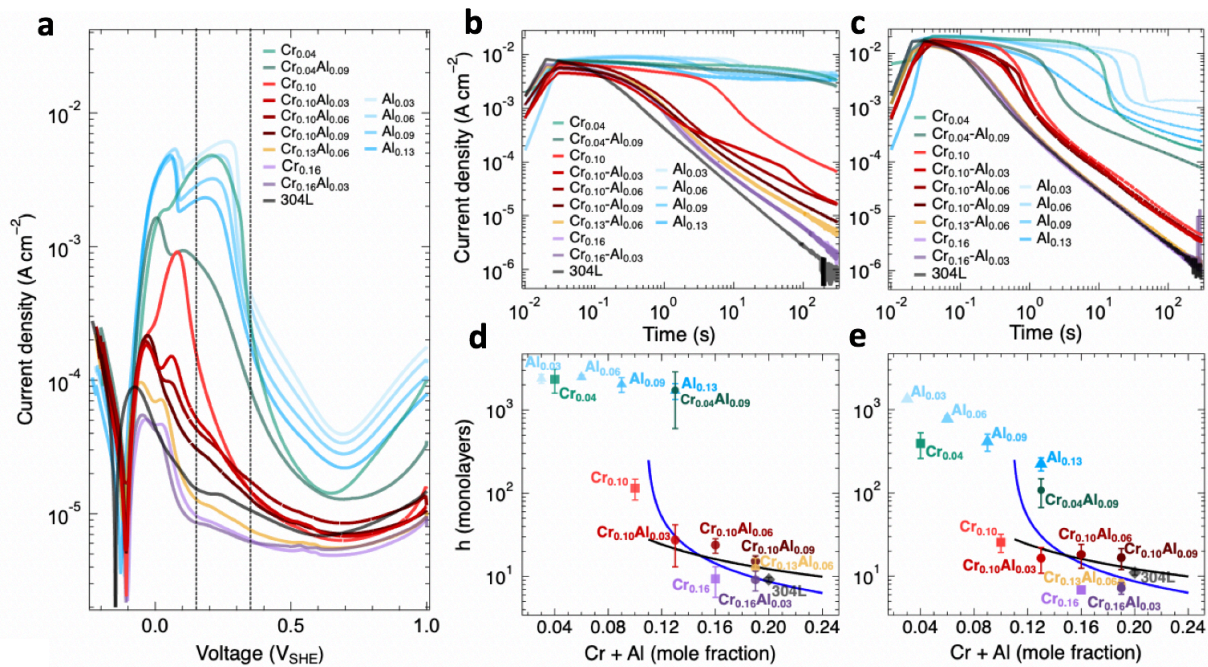

**Fig. 1**. **Electrochemical measures of passive film formation during primary passivation. The alloys are designated in the legends by their Cr$_x$Al$_y$ mole fractions.** (a) LSV behavior for the (FeCoNi)$_{1-x-y}$Cr$_x$Al$_y$ alloys and 304L. The vertical dashed lines at 150 and 350 mV indicate the chronoamperometry voltages. (b) Chronoamperometry for the Cr$_x$Al$_y$ alloys and 304L at 150 mV and (c) 350 mV. (d) *h*- values required for passivation at 150 mV and (e) 350 mV. Error bars in (d) and (e) correspond to the standard deviation in at least 3 data sets.

Since the maximum solubility of Al in all the alloys we examined is only 0.07-0.08 (Supplementary-Fig. S1), the passivation mechanism in Fe-Al binary alloys containing a high-enough mole fraction of Al was likely not operative for these Al containing MPEAs. The LSV of the Cr control samples shows behavior that is similar to that of the Fe-Cr and Ni-Cr binary alloys in sulfuric acid in that as the Cr content increases the passivation wave becomes smaller in amplitude, narrower in width and shifts to lower potentials. In terms of these parameters, the LSV behavior observed for the $(FeCoNi)_{0.9-y}Cr_{0.10}$ + Al alloys was better than the $(FeCoNi)_{0.90}Cr_{0.10}$ control sample.

Figure 1a shows that the corrosion potential of the $(FeCoNi)_{0.9-y}Cr_{0.1}Al_y$ alloys is -110 mV and that of the 304L is -150 mV. The peaks in the corresponding passivation waves also show this 40 mV difference. Following reduction protocols, chronoamperometry was conducted in separate sets of experiments at 150 mV (Fig. 1b) and 350 mV (Fig. 1c) in the pre-passivation regime, termed *primary passivation*, prior to the onset of a steady-state passive current density [13]. In order to assess the role of Al, we compare the required $h$ values for passivation versus the Cr + Al concentrations in the alloys. Numerical integration of chronoamperometry data yields the $h$-values shown in Fig. 1d and Fig.1e which, for comparison, includes results for binary Fe-Cr and Ni-Cr alloys[13]. As is apparent from the LSV, we found that the $h$-values for all the Al-control alloys were large and these alloys did not passivate at potentials up to 350 mV. The $h$-value at 150 mV for $(FeCoNi)_{0.87}Cr_{0.1}Al_{0.03}$ alloy is larger than that of the 304L. We attribute this to the difference in the corrosion and passivation wave peak potentials, since 150 mV is deeper into the primary passivation regime of the 304L. At 350 mV, this difference effectively dissapears, as the $h$-values for the $(FeCoNi)_{0.87}Cr_{0.1}Al_{0.03}$ alloy and the 304L are similar. These results reveal that the MPEA containing 0.10 Cr + 0.03 Al shows passivation behavior significantly better than the 0.10 Cr control alloy and virtually as good as 304L stainless steel containing twice the amount of Cr.

The $(FeCoNi)_{0.81}Cr_{0.10}Al_{0.09}$ alloy has a $h$ value above that for the binary Fe and Ni alloys containing 0.19 Cr because some of the Al in the FCC phase was taken up by the formation of a $\beta$-NiAl phase which occupied 3.5% of the alloy surface ( Supplementary Fig. S7-S8). All the $(FeCoNi)_{0.9-y}Cr_{0.10}Al_y$ alloys, except the 0.09Al alloy display lower $h$-values at 350 mV compared to that at 150 mV. Within experimental scatter, the 0.09Al alloy shows the same $h$-value at both voltages as a result of $\beta$-NiAl phase dissolution (Supplementary Fig. S9). This suggests that dissolution of this phase is the reason that the $h$-values are larger than those of the binary Fe-Cr and Ni-Cr alloys.

Figure 2a shows the electrochemical impedance spectroscopy (EIS) results obtained for the MPEAs in the primary passivation regime at 150 mV. This measure is indicative of the polarization resistance, which was best for the $Cr_{0.13}Al_{0.06}$ and the $Cr_{0.16}Al_{0.03}$ alloys. Figures 2b and 2c show representative XPS spectra for the $Cr^{3+}$ $2p_{3/2}$ [24] and $Al^{3+}$ 2p [25,26] core shells characterizing the composition of the passive film formed on the $(FeCoNi)_{0.87}Cr_{0.10}Al_{0.03}$ alloy and Supplementary Fig. S10-S13 shows the XPS results obtained for the full set of $(FeCoNi)_{0.9-y}Cr_{0.10}Al_y$ alloys. Fig. 2d is a parametric plot showing that the polarization resistance increases with the $Cr^{3+}$ + $Al^{3+}$ content of the passive film.

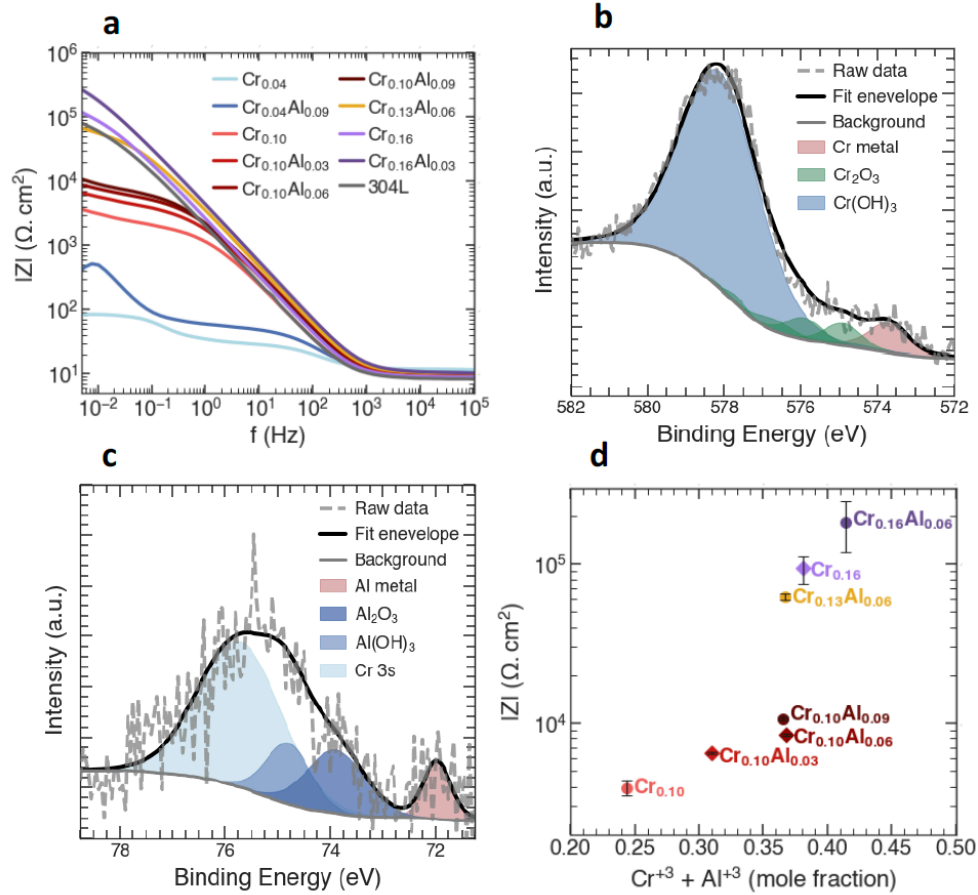

**Fig. 2. EIS and XPS characterization of the passive film formed during primary passivation.** (a) EIS Bode plot for (FeCoNi)$_{1-x-y}$Cr$_x$Al$_y$ alloys for x ≥ 0.10 and 304L stainless steel following a potentiostatic hold at 150 mV for 10 ks. (b) XPS Cr2p$_{3/2}$ and (c) Al2p core shell binding energy peak deconvolutions for the determination of Cr$^{3+}$ and Al$^{3+}$ mole fraction in the passive film formed on the (FeCoNi)$_{0.87}$Cr$_{0.10}$Al$_{0.023}$ alloy. (d) Parametric plot for |Z| at 5 mHz, as a function of the Cr$^{3+}$ and Al$^{3+}$ mole fraction in the passive film.

### 3.2 *Neutron Scattering Measurements of Warren-Cowley CSRO Parameters*

Initial Rietveld refinements of the neutron Bragg diffraction data confirmed that the phases were FCC [27,28]. The ratio of intensities of the (200) and (111) Bragg reflections was observed to vary between samples, indicative of either preferred orientation or short-range order. Preferred orientation is excluded as an explanation, as the ratio of the (400) and (222) Bragg reflections behaved different from that of the (200)/(111) pair. This indicates that short-range order must be present in these MPEAs. Furthermore, the highest Al content phase, (FeCoNi)$_{0.81}$Cr$_{0.10}$Al$_{0.09}$, also contained minor peaks indicative of a 46 at% Al, β-NiAl phase. PDF refinements were performed using PDFgui and PDFfit2 [28,29]. The PDF data sets were fit using $D(r) = 4\pi r \rho_o G(r)$, where $G(r)$ is the total radial distribution function and $\rho_o$ is equal to the total number of atoms per unit volume [30].

Reverse Monte Carlo (RMC) refinements of the PDF data were performed using 20 x 20 x 20 FCC unit cells containing 32,000 atoms. The CSRO was determined for the (FeCoNi)$_{0.9-y}$Cr$_{0.10}$Al$_y$ alloys[30-32]. Representative images and fits of the PDF data for (FeCoNi)$_{0.9-y}$Cr$_{0.10}$Al$_y$ single-phase alloys are shown in Fig. 3a and Fig. 3b. The average number per unit volume of Cr atoms that form Cr-Cr pairs was determined from the RMC results and is shown in Fig. 3c.

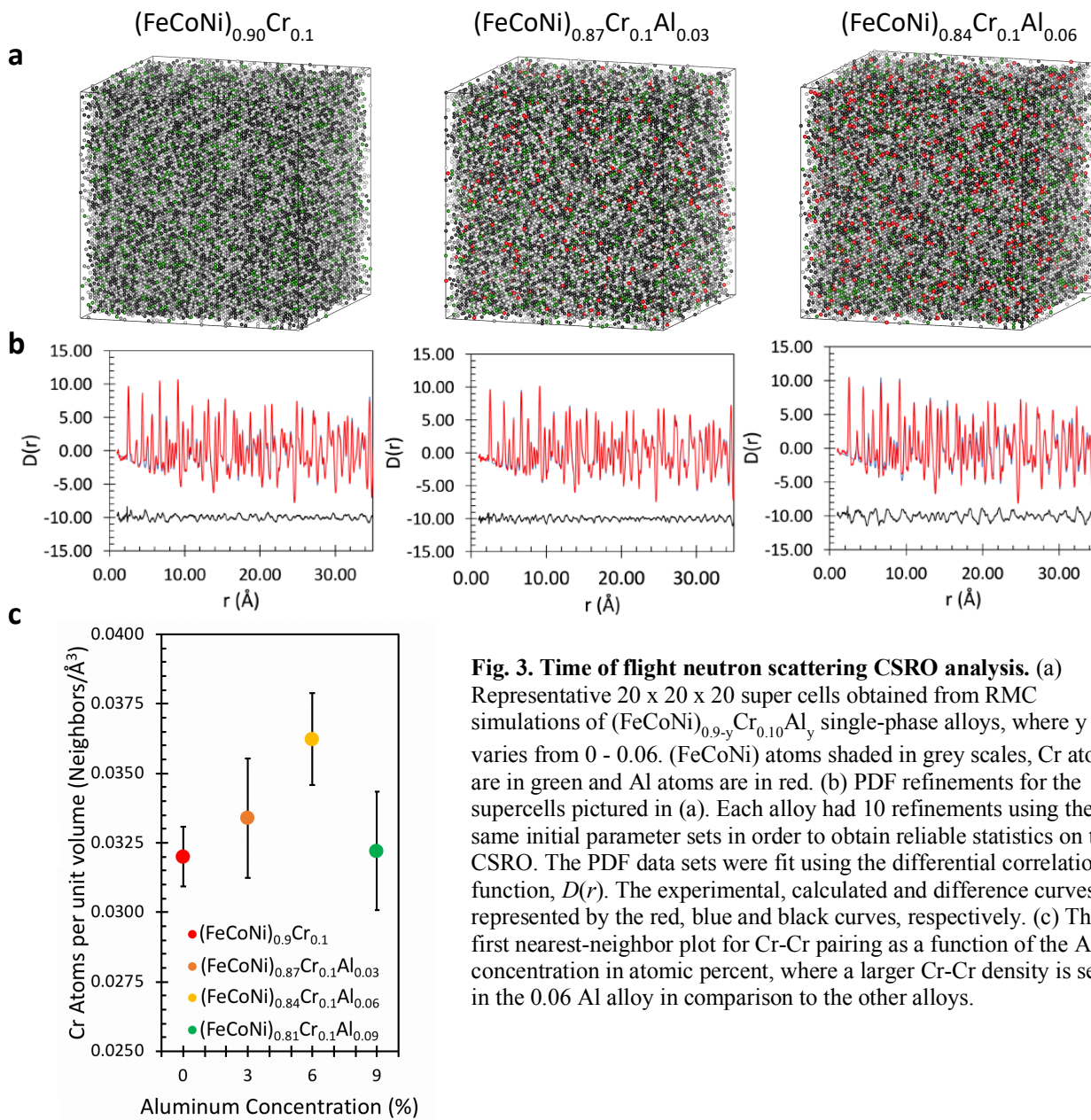

**Fig. 3. Time of flight neutron scattering CSRO analysis.** (a) Representative 20 x 20 x 20 super cells obtained from RMC simulations of $(FeCoNi)_{0.9-y}Cr_{0.10}Al_y$ single-phase alloys, where y varies from 0 - 0.06. (FeCoNi) atoms shaded in grey scales, Cr atoms are in green and Al atoms are in red. (b) PDF refinements for the supercells pictured in (a). Each alloy had 10 refinements using the same initial parameter sets in order to obtain reliable statistics on the CSRO. The PDF data sets were fit using the differential correlation function, $D(r)$. The experimental, calculated and difference curves are represented by the red, blue and black curves, respectively. (c) The first nearest-neighbor plot for Cr-Cr pairing as a function of the Al concentration in atomic percent, where a larger Cr-Cr density is seen in the 0.06 Al alloy in comparison to the other alloys.

There is a notable increase in the density of Cr-Cr pairs as the Al content increases, correlating with the passivation behavior of the $(FeCoNi)_{1-x-y}Cr_xAl_y$ alloys. The density of such pairs decreases for the $(FeCoNi)_{0.81}Cr_{0.1}Al_{0.09}$ alloy, that we attribute to the reduced Al within the primary FCC phase as a result of the limited solubility of Al and the formation of β-NiAl

The multi-component generalization of Warren-Cowley (WC) short range order parameters were computed and are reported in Table 1 for Cr-Cr, Co-Al and Ni-Al pairs. Supplemental Information Table S4, shows a more complete list. These parameters, are defined as $\alpha_{ii}^{p} = \frac{P_{ii}(p) - \bar{c}_i}{1 - \bar{c}_i}$ and $\alpha_{ij}^{p} = 1 - \frac{P_{ij}(p)}{\bar{c}_j}$ $(i \neq j)$ [43]. Here $P_{ij}(p)$ is the probability of finding an atom of type $j$, at $(p + r)$ given an atom of type $i$ at $(r)$ and $\bar{c}_i$ or $\bar{c}_j$ is the average concentration of the species in the lattice. For a central $i$-atom located at $r$, $p$ corresponds to the distances defined by the 1st, 2nd, 3rd … nth coordination shells. For $i \neq j$, $\alpha_{ij}^{p} < 0$, corresponds to the "ordering" type of CSRO. For example, in a FCC crystal lattice which has 12 first nearest neighbors, for $\alpha < 0$, it is energetically more favorable for a central $i$-atom to have statistically more $j$ atoms than would occur if there was ideal mixing of the components for which $\alpha = 0$ $(P_{ij}(p) = \bar{c}_j)$. If $\alpha_{ij}^{p} > 0$, the opposite is true, resulting in a tendency for the components to cluster. $\alpha_{ii}^{p}$ is the Warren-Cowley SRO parameter for like elemental pairs and has the same sign for ordering or clustering as $\alpha_{ij}^{p}$.

It is important to note that this neutron method formally measures the CSRO averaged over the entire volume of the sample so the results directly reflect the average CSRO present in the sample, and the uncertainties in the CSRO values describe the range of CSRO over which almost all (in a statistical sense) individual regions exist. The primary passivation process that we electrochemically characterized senses the same CSRO in the alloy as that deduced from the neutron scattering.

**Table 1.** Selected CSRO parameters determined from time of flight neutron scattering data.

$\alpha_{Cr-Cr}$

| Cr-Cr | $(FeCoNi)_{0.9}Cr_{0.1}$ | $(FeCoNi)_{0.87}Cr_{0.1}Al_{0.03}$ | $(FeCoNi)_{0.84}Cr_{0.1}Al_{0.06}$ | $(FeCoNi)_{0.81}Cr_{0.1}Al_{0.09}$ |
|---|---|---|---|---|
| 1st NN | $0.010 \pm 0.002$ | $0.016 \pm 0.003$ | $0.020 \pm 0.002$ | $0.018 \pm 0.002$ |
| 2nd NN | $-0.005 \pm 0.003$ | $0.011 \pm 0.004$ | $0.020 \pm 0.003$ | $0.011 \pm 0.002$ |
| 3rd NN | $0.001 \pm 0.002$ | $0.005 \pm 0.001$ | $0.015 \pm 0.002$ | $0.009 \pm 0.002$ |
| 4th NN | $-0.001 \pm 0.003$ | $0.002 \pm 0.002$ | $0.006 \pm 0.003$ | $0.005 \pm 0.003$ |

$\alpha_{Co-Al}$

| Co-Al | $(FeCoNi)_{0.9}Cr_{0.1}$ | $(FeCoNi)_{0.87}Cr_{0.1}Al_{0.03}$ | $(FeCoNi)_{0.84}Cr_{0.1}Al_{0.06}$ | $(FeCoNi)_{0.81}Cr_{0.1}Al_{0.09}$ |
|---|---|---|---|---|
| 1st NN | - | $-0.20 \pm 0.03$ | $-0.224 \pm 0.009$ | $-0.238 \pm 0.006$ |
| 2nd NN | - | $-0.15 \pm 0.02$ | $-0.24 \pm 0.01$ | $-0.16 \pm 0.01$ |
| 3rd NN | - | $-0.06 \pm 0.01$ | $-0.18 \pm 0.01$ | $-0.134 \pm 0.004$ |
| 4th NN | - | $-0.02 \pm 0.02$ | $-0.08 \pm 0.01$ | $-0.08 \pm 0.01$ |

$\alpha_{Ni-Al}$

| Ni-Al | $(FeCoNi)_{0.9}Cr_{0.1}$ | $(FeCoNi)_{0.87}Cr_{0.1}Al_{0.03}$ | $(FeCoNi)_{0.84}Cr_{0.1}Al_{0.06}$ | $(FeCoNi)_{0.81}Cr_{0.1}Al_{0.09}$ |
|---|---|---|---|---|
| 1st NN | - | $0.15 \pm 0.02$ | $0.193 \pm 0.009$ | $0.21 \pm 0.01$ |
| 2nd NN | - | $0.10 \pm 0.02$ | $0.20 \pm 0.01$ | $0.15 \pm 0.01$ |
| 3rd NN | - | $0.04 \pm 0.01$ | $0.149 \pm 0.009$ | $0.120 \pm 0.005$ |
| 4th NN | - | $0.02 \pm 0.02$ | $0.06 \pm 0.02$ | $0.068 \pm 0.007$ |

## 4.0 Computational Results

To help interpret the experimental CSRO results, we computed the composition-dependent SRO parameters at elevated temperature using a combination of DFT [33-37], cluster expansion [38-41] and Monte Carlo calculations [42]. For computational simplicity, the full multicomponent $(FeCoNi)_{0.9-y}Cr_{0.10}Al_y$ alloys were reduced to the representative ternary alloys, $Ni_{0.9-y}Cr_{0.10}Al_y$, and $Co_{0.9-y}Cr_{0.10}Al_y$, as we are interested in the properties of a transition metal with alloying additions of Al and Cr. CSRO calculations were obtained for a range of ternary compositions by equilibrating the disordered alloys and averaging the pair-wise correlations. A temperature of T=3000 K (Ni alloy) and 3600 K (Co alloy) was necessary to ensure that the alloys were maintained in the single-phase region. We note that this overestimation of phase boundary temperatures is well documented for cluster expansions [38-41]. The W-C CSRO parameters for Cr-Cr, Al-Ni and Al-Cr pairs in the ternary diagrams are shown in Figs. 4a- 4d. Figures 4e and 4f display the detailed changes in the six CSRO parameters for the alloys.

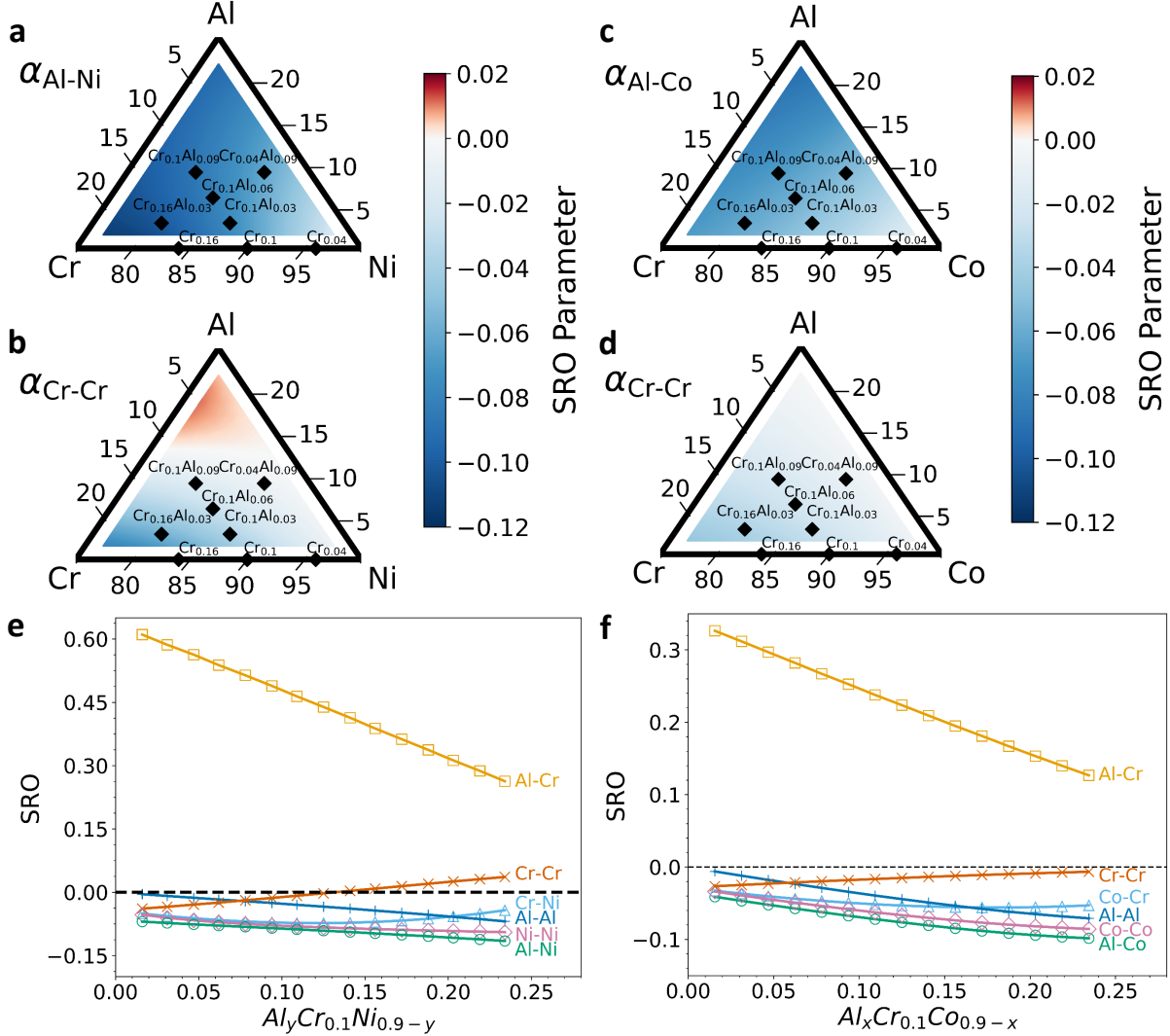

**Fig. 4. First principles cluster expansion and Monte Carlo results for Ni$_{0.9-y}$Cr$_{0.10}$Al$_y$ and Co$_{0.9-y}$Cr$_{0.10}$Al$_y$ alloys.** Ternary diagrams of (a) Al-Ni SRO and (b) Cr-Cr SRO in the Ni-rich portion of the Al-Cr-Ni phase space. The adjacent color bar indicates the CSRO parameters. (c) Al-Co SRO and (d) Cr-Cr SRO in the Co-rich portion of the Al-Cr-Co phase space. The adjacent color bars indicate the CSRO parameters. Experimental compositions are overlaid in black diamonds. SRO parameter vs. Al mole fraction for (e) Ni$_{0.9-y}$Cr$_{0.10}$Al$_y$ alloys and (f) Co$_{0.9-y}$Cr$_{0.10}$Al$_y$ alloys.

The full CSRO results for all pairs in the Ni and Co ternary alloys are shown in Supplementary Figs. S14-S15. It is evident that both sets of alloys exhibit a pronounced increase in ordering of Al-Al pairs as the Al-content is increased, whereas, the CSRO parameter of Cr-Cr pairs tends toward clustering with increased Al. For both alloys, the Al-Cr pairs are strongly clustering, but with rapidly decreasing magnitude with increasing Al content. We point out that a direct quantitative comparison of the magnitude of the Al-Cr CSRO parameters in the ternary alloys is not justified as a result of the temperatures used to obtain single phases in the cluster expansions. However, in each case the qualitative trends are the same; an increase in Cr-Cr and decrease in Al-Al and Al-Cr CSRO parameters is driven by the strong ordering tendency of Al-Ni and Al-Co pairs.

Depending on the composition, both like-species pairs (Cr-Cr or Al-Al) can either cluster or order in the ternarys. The W-C $\alpha_{Cr\text{-}Cr}$ shows the ordering type of CSRO with no Al, but its clustering tendency increases as the Al concentration increases. Conversely, $\alpha_{Al\text{-}Al}$ orders over this whole range in composition, with the strength of the tendency increasing with additional Al. Neither ordering tendency is unexpected. Figures 4a and 4c show that the Al-Ni, Cr-Ni and Al-Co ground state phase diagrams exhibit ordered compounds, as reflected in the CSRO tendency becoming stronger when moving closer to each respective binary edge in the ternary phase diagram. However, the increase in clustering tendency of Cr-Cr pairs with increasing Al concentration, while consistent with the neutron scattering measurements, is surprising given the ordering tendencies in the binary phase diagrams. We have investigated this tendency in more detail using Monte Carlo simulations of a variety of types of ternary systems. We find that the Cr-Cr clustering is enhanced because of the strong ordering tendency between Ni or Co and Al, leading to the formation of a preference for Ni-Al or Co-Al pairs and effectively "starving" the Cr atoms of Ni and Al atoms with which to form pairs. In essence, Al-Ni ordering deprives Al from interacting with Cr.

The calculated SRO parameters for Al-Cr in the ternary alloys (Figs. 4e and 4f) have a large clustering tendency. The neutron scattering data for the (FeCoNi)$_{0.9\text{-}y}$Cr$_{0.10}$Al$_y$ alloys shows that there is a strong "ordering-type" (negative values) W-C parameter between Al and Co nearest-neighbors across all compositions (Table I), in agreement with our DFT result. The neutron data also shows weaker "clustering-type" W-C parameters between Fe-Al and Ni-Al (Supplemental Table S4). The strong Co-Al ordering tendency means that each Al atom prefers to have a Co atom as a neighbor relative to the other elements present. This results in fewer Ni atom neighbors per Al than expected from random sampling, so the Ni-Al WC parameters becomes "clustering-type" (positive). This effect is accentuated by the large imbalance in concentrations of Co and Al so the relative surplus of Co further suppresses Ni-Al pairs. These complicated interactions in the multicomponent alloys can not occur in the model ternary alloys examined.

## 5. Percolation passivation applied to MPEAs

The percolation theory of passivity [13] is based on the concept that a system of finite thickness, $h$, can be renormalized to a two-dimensional (2D) system by changing the 2D site percolation threshold, $p_c^{2D}$, to a series of thickness-dependent percolation thresholds, $p_c(h)$ [44,45]. In our experiments, $h$ represents the number of atomic layers that have to undergo selective dissolution of Fe, Co and Ni in order for metallic Cr atoms to percolate across the corrosion roughened surface. Theory predicts that $h = c[p_c(h) - p_c^{3D}]^{-\nu_{3D}}$, where $p_c^{3D}$ is the relevant 3D site percolation threshold, $c$ is a constant of order unity and $\nu_{3D}$ is the percolation correlation length scaling exponent which has a universal value of 0.878 in 3D [13]. For the case of the (FeCoNi)$_{1\text{-}x\text{-}y}$Cr$_x$Al$_y$ alloys, the relevant $p_c^{3D}$ is determined by the spanning distance of the atomic radii of Cr$^{3+}$ and O$^{2\text{-}}$ ions [46] and nearest neighbor distances in the FCC lattice. The maximum spanning distance is about 0.40 nm which is between the 2$^{nd}$ and 3$^{rd}$ nearest-neighbor distance in the (FeCoNi)$_{1-y}$Cr$_{0.10}$Al$_y$ lattices (Supplementary Fig. S16). This separation distance sets the minimum mole fraction of Cr in the lattice that allows for the development of percolating Cr oxide/oxyhydroxide clusters. The relevant value of $p_c^{3D}$ is the percolation threshold for $1^{st} + 2^{nd}$ nearest neighbors, which for ideal mixing is 0.136 [47]. However, the neutron scattering data shows that the alloys are ordered, which results in a decrease of $p_c^{3D}$ [13]. We performed a best fit analysis for $p_c^{3D}$ based on the experimentally obtained $h$-values at 150 mV for the (FeCoNi)$_{0.87}$Cr$_{0.10}$Al$_{0.03}$, (FeCoNi)$_{0.81}$Cr$_{0.13}$Al$_{0.06}$ and the (FeCoNi)$_{0.81}$Cr$_{0.16}$Al$_{0.03}$ alloys and found a value equal to 0.077. The magnitude of the reduction of $p_c^{3D}$ from the ideal mixing case for the MPEAs is similar to that for the clustering type of SRO in BCC lattices [13] as well as FCC [48] and simple cubic lattices [49]. It is important to note that it is the clustering type of CSRO that significantly lowers the site percolation threshold, thus reducing the minimum mole fraction of Cr required for percolation across the sample surface and concomitantly the minimum Cr content for excellent passivation [13]. The model prediction for the $h$-values is compared to the experimental $h$-data for these alloys in Supplementary Fig. S17.

## 6. Outlook and Concluding Remarks

Our results reveal that the addition of only 0.03 - 0.06 Al to a (FeCoNi)$_{0.90}$Cr$_{0.10}$ MPEA significantly benefits passivation behaviors. The time-of-flight neutron scattering study found that the Al addition increased the Cr-Cr clustering type of CSRO which results in a reduced site percolation threshold of the Cr component and an accompanying improvement in passivation behavior. An interesting question is whether further reductions in the required Cr content is possible using this strategy, while still retaining excellent passivation behavior. Based on our current knowledge, this should be possible by adding another element to the alloy that further increases the Cr-Cr CSRO parameter. Importantly, this general approach provides a new design path for improving the passivation behavior by increasing the degree of CSRO-clustering of the passivating component(s) comprising an alloy.

**Acknowledgments**

MT, JRS, CW, TMM and KS acknowledge support of this research by the Office of Naval Research, Multidisciplinary University Research Initiative program, "From Percolation to Passivation (P2P): Multiscale Prediction and Interrogation of Surface and Oxidation Phenomena in Multi-Principal Element Alloys" under Grant Number N00014-20-1-2368. The neutron scattering portion of this research conducted at Oak Ridge National Laboratory's Spallation Neutron Source was sponsored by the Scientific User Facilities Division, Office of Basic Energy Sciences, U.S. Department of Energy. The neutron analysis was performed using the Advanced Research Computing at Hopkins (ARCH) core facility (rockfish.jhu.edu), which is supported by the National Science Foundation under grant number OAC 1920103. Materials processing was performed at the Platform for the Accelerated Realization, Analysis, and Discovery of Interface Materials (PARADIM at JHU), a National Science Foundation Materials Innovation Platform under grant number NSF DMR-2039380.

**Competing Interests**

The authors declare no competing interests.

**Data availability**

All the data from this work is available upon reasonable request from the corresponding author.

**Additional Information**

Supplementary Information: Supplementary Figs. S1-S17 and Tables S1-S4.
Correspondence and requests for materials should be addressed to K.S.

**Figure Captions**

Fig. 1. Electrochemical measures of passive film formation during primary passivation. The alloys are designated in the legends by their $Cr_xAl_y$ mole fractions. (a) LSV behavior for the $(FeCoNi)_{1-x-y}Cr_xAl_y$ alloys and 304L. The vertical dashed lines at 150 and 350 mV indicate the chronoamperometry voltages. (b) Chronoamperometry for the $Cr_xAl_y$ alloys and 304L at 150 mV and (c) 350 mV. (d) $h$- values required for passivation at 150 mV and (e) 350 mV. Error bars in (d) and (e) correspond to the standard deviation in at least 3 data sets.

Fig. 2. EIS and XPS characterization of the passive film formed during primary passivation. (a) EIS Bode plot for $(FeCoNi)_{1-x-y}Cr_xAl_y$ alloys for $x \geq 0.10$ and 304L stainless steel following a potentiostatic hold at 150 mV for 10 ks.

(b) XPS Cr2p$_{3/2}$ and (c) Al2p core shell binding energy peak deconvolutions for the determination of Cr$^{3+}$ and Al$^{3+}$ mole fraction in the passive film formed on the (FeCoNi)$_{0.87}$Cr$_{0.10}$Al$_{0.023}$ alloy. (d) Parametric plot for $|Z|$ at 5 mHz, as a function of the Cr$^{3+}$ and Al$^{3+}$ mole fraction in the passive film.

Fig. 3. Time of flight neutron scattering CSRO analysis. (a) Representative 20 x 20 x 20 super cells obtained from RMC simulations of (FeCoNi)$_{0.90\text{-}y}$Cr$_{0.10}$Al$_y$ single-phase alloys, where y varies from 0 - 0.06. (FeCoNi) atoms shaded in grey scales, Cr atoms are in green and Al atoms are in red. (b) PDF refinements for the supercells pictured in (a). Each alloy had 10 refinements using the same initial parameter sets in order to obtain reliable statistics on the CSRO. The PDF data sets were fit using the differential correlation function, $D(r)$. The experimental, calculated and difference curves are represented by the red, blue and black curves, respectively. (c) The first nearest-neighbor plot for Cr-Cr pairing as a function of the Al concentration in atomic percent, where a larger Cr-Cr density is seen in the 0.06 Al alloy in comparison to the other alloys.

Fig. 4. First principles cluster expansion and Monte Carlo results for Ni$_{0.9\text{-}y}$Cr$_{0.10}$Al$_y$ and Co$_{0.9\text{-}y}$Cr$_{0.10}$Al$_y$ alloys. Ternary diagrams of (a) Al-Ni SRO and (b) Cr-Cr SRO in the Ni-rich portion of the Al-Cr-Ni phase space. The adjacent color bar indicates the CSRO parameters. (c) Al-Co SRO and (d) Cr-Cr SRO in the Co-rich portion of the Al-Cr-Co phase space. The adjacent color bars indicate the CSRO parameters. Experimental compositions are overlaid in black diamonds. SRO parameter vs. Al mole fraction in (e) Ni$_{0.9\text{-}y}$Cr$_{0.10}$Al$_y$ and (f) Co$_{0.9\text{-}y}$Cr$_{0.10}$Al$_y$

**SUPPLEMENTARY MATERIAL**

# Tuning chemical short-range order for stainless behavior at reduced chromium concentrations in multi-principal element alloys

W.H. Blades[1], B.W.Y. Redemann[2,3,4], N. Smith[5], D. Sur[6], M.S. Barbieri[6], Y. Xie[1], S. Lech[3], E. Anber[3], M.L. Taheri[3], C. Wolverton[5], T.M. McQueen[2,3,4], J.R. Scully[6], K. Sieradzki[1]*

[1]*Ira A. Fulton School of Engineering; Arizona State University; Tempe, AZ, 85287, USA.*

[2]*Department of Chemistry, The Johns Hopkins University; Baltimore, MD, 21218, USA.*

[3]*Department of Material Science and Engineering, The Johns Hopkins University; Baltimore, MD, 21218, USA.*

[4]*William H. Miller III Department of Physics and Astronomy, The Johns Hopkins University; Baltimore, MD, 21218, USA.*

[5]*Department of Materials Science and Engineering, Northwestern University; Evanston, IL, 60208, USA*

[6]*Department of Material Science and Engineering, University of Virginia; Charlottesville, VA, 22904, USA.*

*karl.sieradzki@asu.edu

This Supplementary Material file includes:

Figs. S1 - S17
Tables S1 - S4

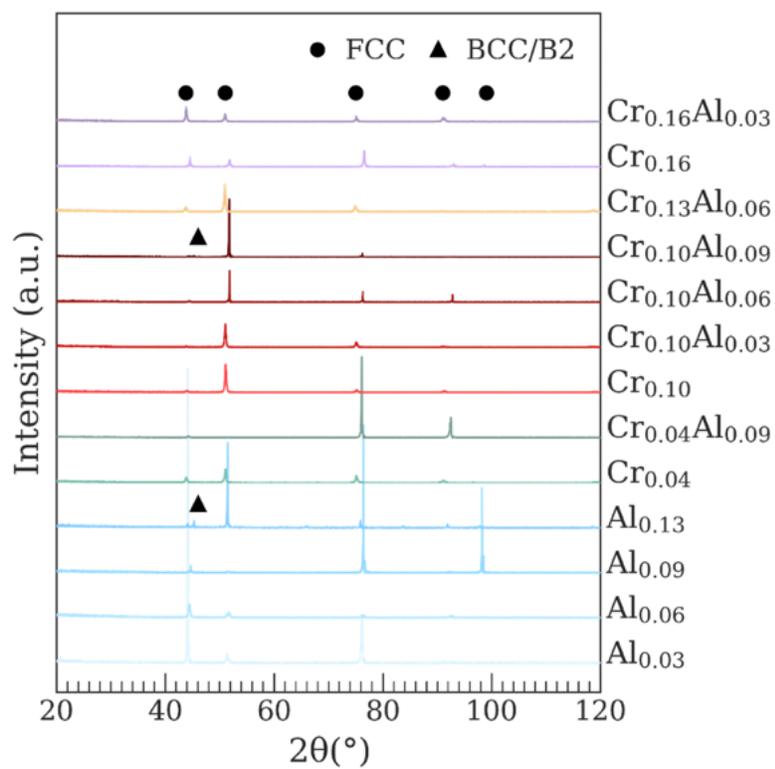

**Figure S1. X-ray diffraction patterns of the (FeCoNi)$_{1-x-y}$Cr$_x$Al$_y$ MPEAs.** Samples were homogenized at 1000 °C and quenched in ice water.

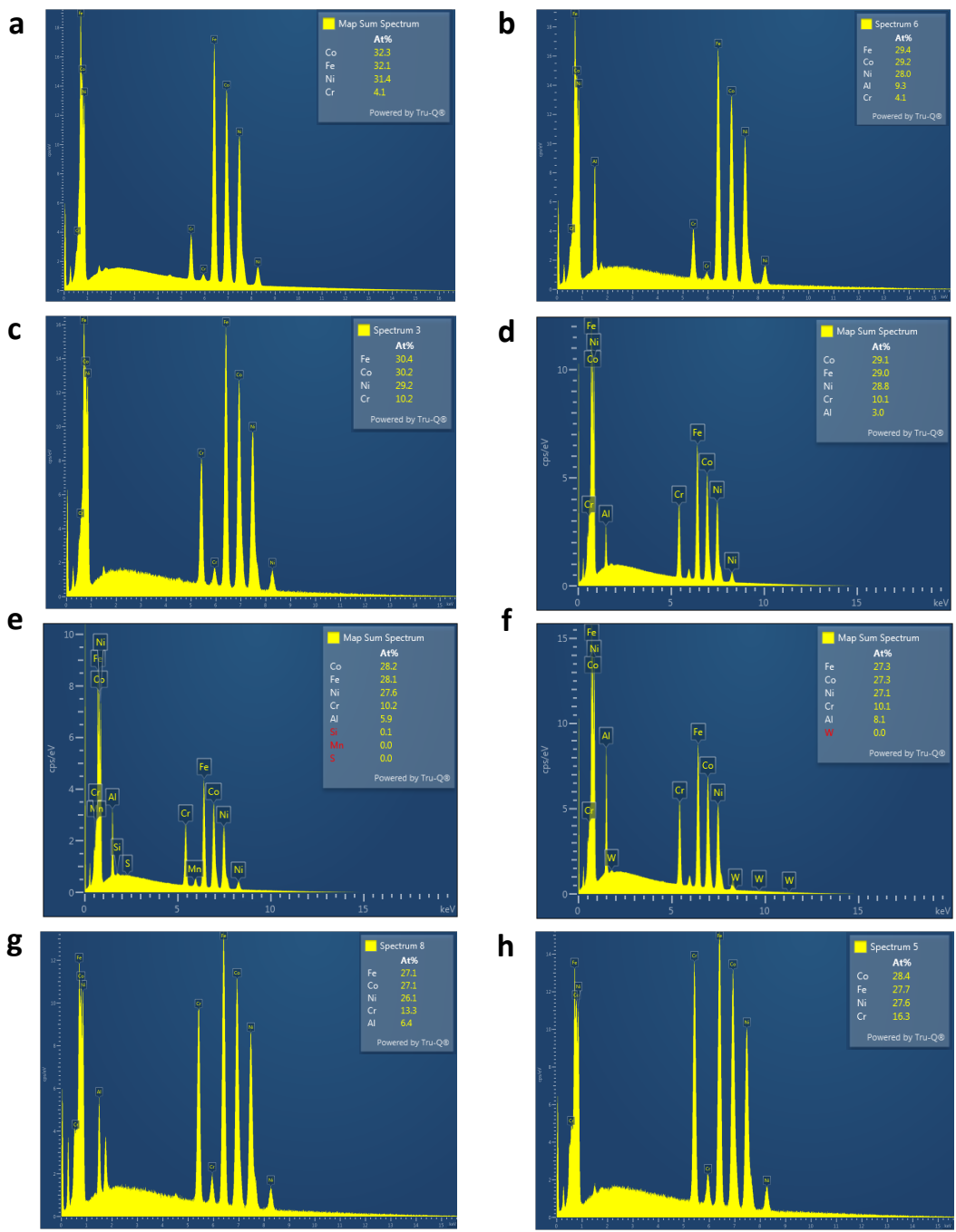

**Figure S2. SEM/EDS composition characterization of the (FeCoNi)$_{1-x-y}$Cr$_x$Al$_y$ of the MPEAs.**
**(a)** (FeCoNi)$_{0.96}$Cr$_{0.04}$ **(b)** (FeCoNi)$_{0.87}$Cr$_{0.04}$Al$_{0.09}$ **(c)** (FeCoNi)$_{0.90}$Cr$_{0.10}$ **(d)** (FeCoNi)$_{0.87}$Cr$_{0.10}$Al$_{0.03}$
**(e)** (FeCoNi)$_{0.84}$Cr$_{0.10}$Al$_{0.06}$ **(f)** (FeCoNi)$_{0.81}$Cr$_{0.10}$Al$_{0.09}$ **(g)** (FeCoNi)$_{0.81}$Cr$_{0.13}$Al$_{0.06}$ **(h)** (FeCoNi)$_{0.84}$Cr$_{0.16}$

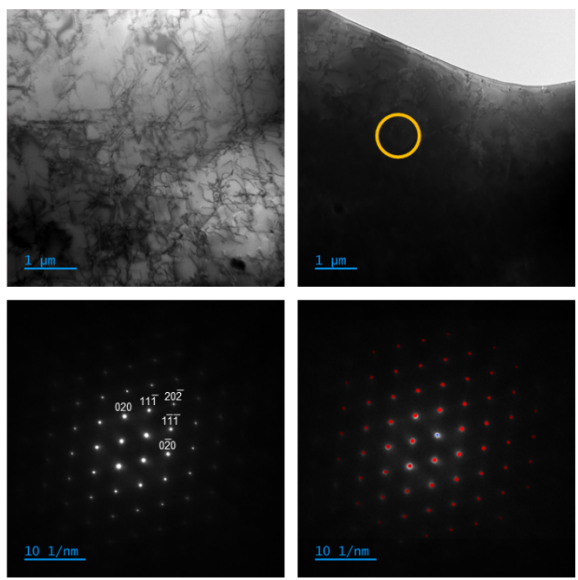

**Figure S3. An example of TEM conducted on the MPEAs: [(FeCoNi)$_{0.87}$Cr$_{0.10}$Al$_{0.03}$] in a bright field mode and a crystal structure analysis.** The yellow circle corresponds to the selective aperture position during the diffraction pattern acquisition. The bottom row shows the acquired SAED pattern overlapped with an fcc-Ni pattern calculated in JEMS software. The lattice parameter is 0.36 nm.

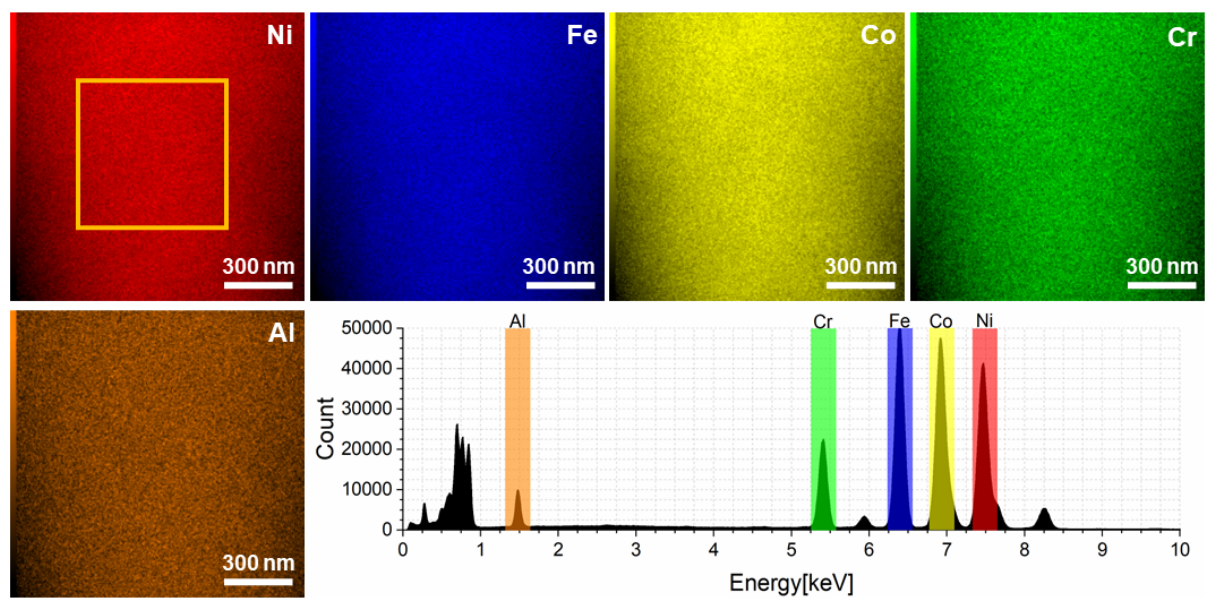

**Figure. S4. Example of STEM-EDS analysis for the (FeCoNi)$_{0.87}$Cr$_{0.10}$Al$_{0.03}$ alloy.** The STEM maps confirm the uniform distribution of the elements. The EDS spectrum corresponds to the area marked with a yellow rectangle. The thickness corrected composition is (FeCoNi)$_{0.867}$Cr$_{0.10}$Al$_{0.0282}$

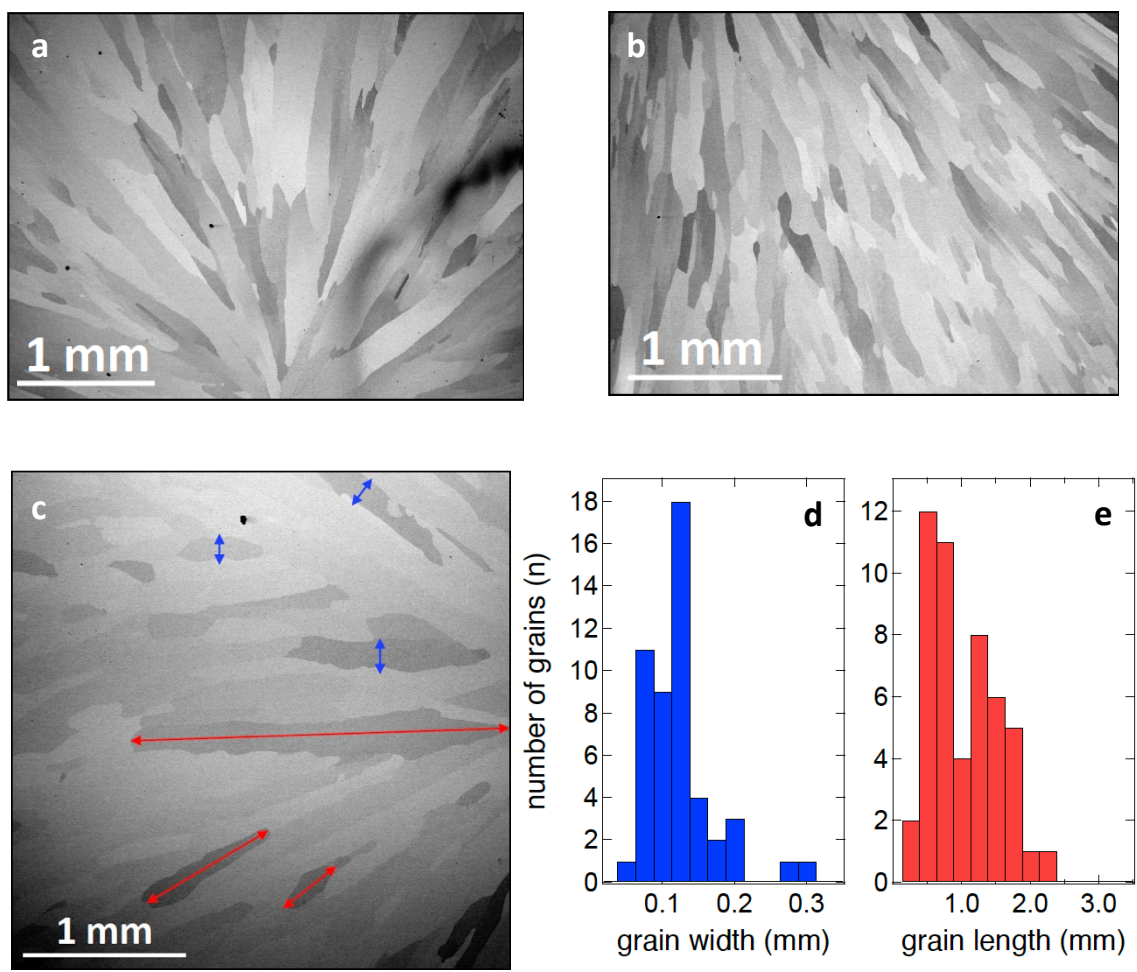

**Figure S5.** Grain size analysis of the of the (FeCoNi)$_{0.90}$Cr$_{0.10}$Al$_y$ sample. SEM images **(a)** and **(b)** show the large columnar grain structure of the sample. (**c**) Blue and red arrows superimposed over the image highlight how the grain width **(d)** and grain length **(e)** were determined.

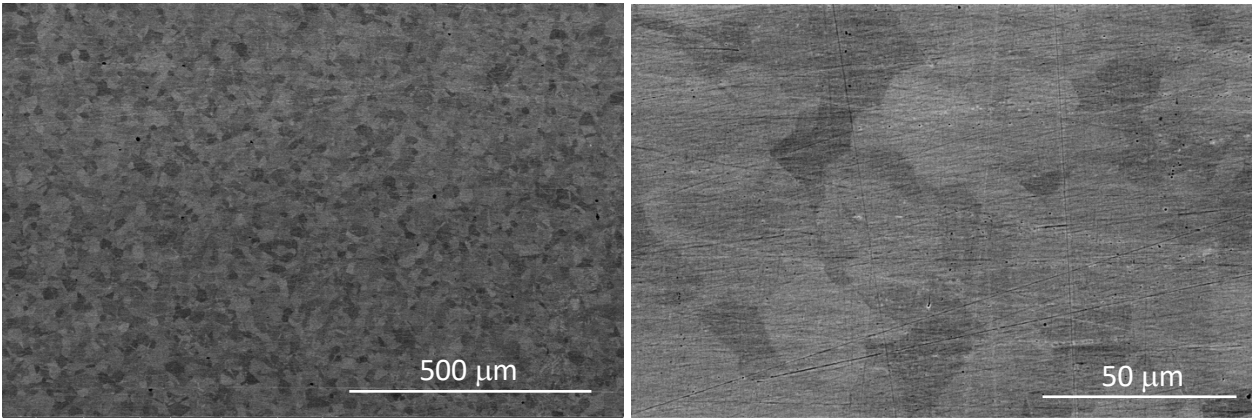

**Fig. S6. 304 L stainless steel microstructure.** Grain shapes vary throughout the microstructure but are primarily equiaxed. The average grain size is 40 μm.

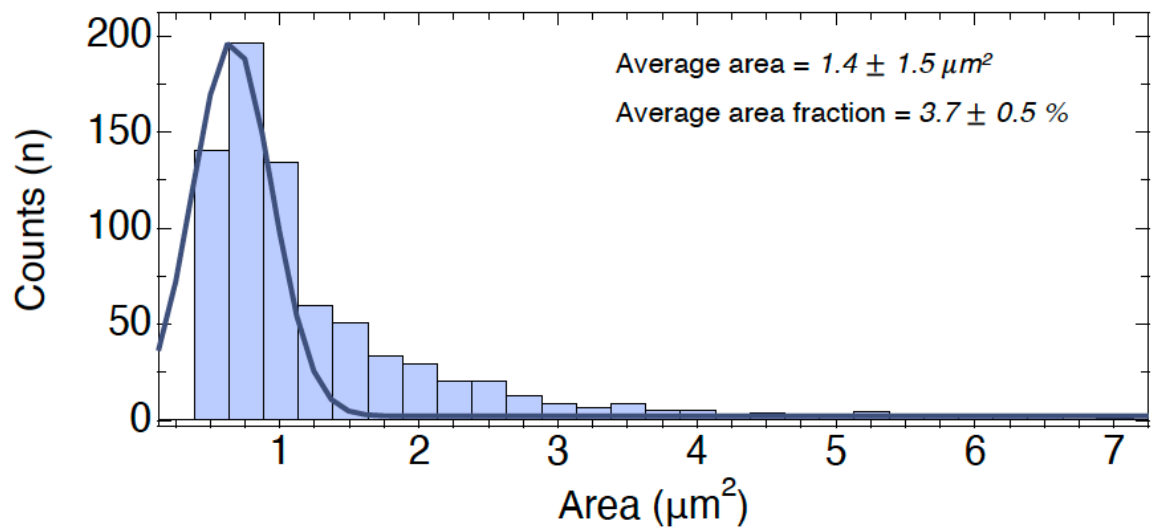

**Figure S7. Area histogram of the second phase regions measured from the SEM micrograph segmentation analysis on the (FeCoNi)$_{0.81}$Cr$_{0.1}$Al$_{0.09}$ sample surface.** The lighter blue bars represent the number of second phase area counts, binned at 0.25 $\mu$m$^2$ intervals, and the darker line superimposed over the bars is a Gaussian fit to the data.

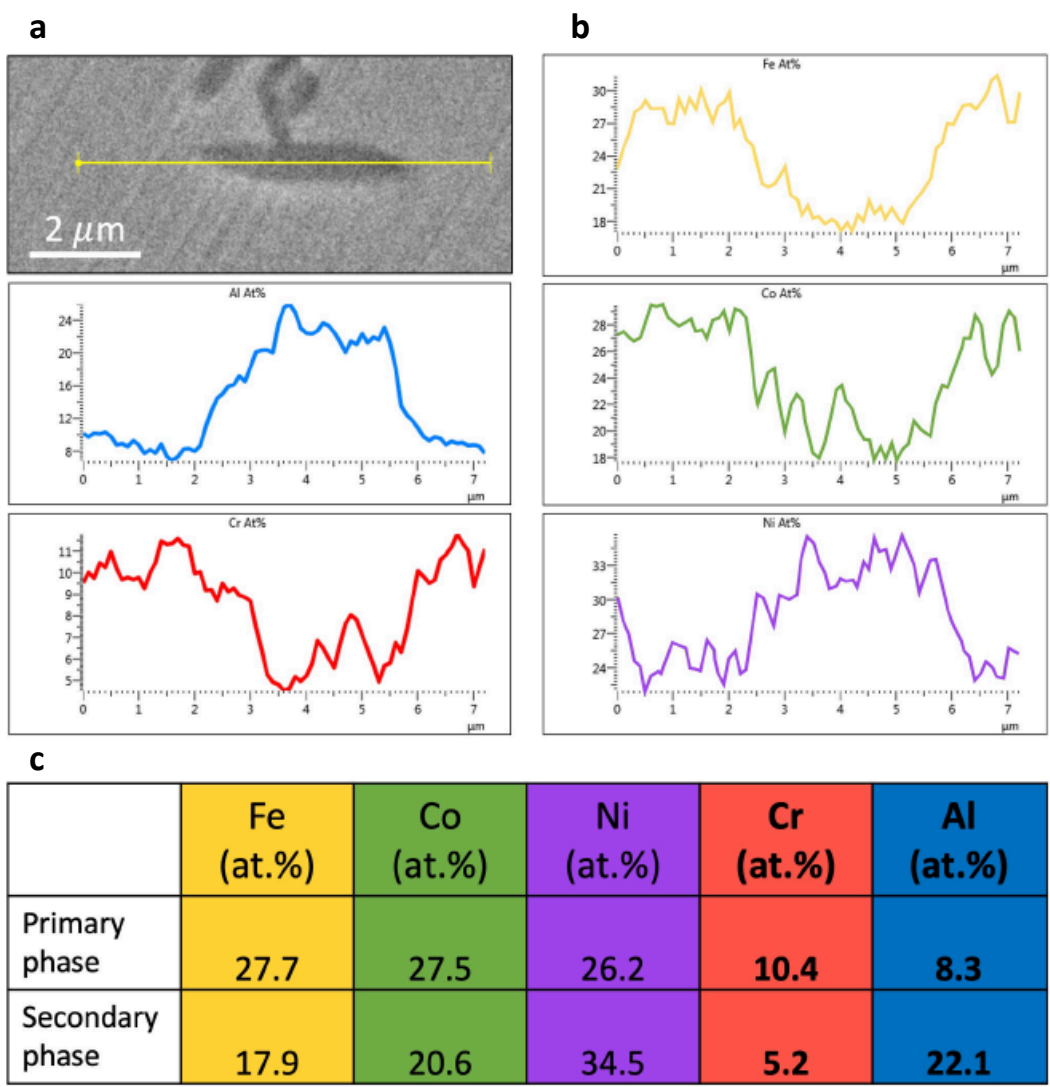

**Figure S8. SEM image and EDS measurement on the (FeCoNi)$_{0.81}$Cr$_{0.1}$Al$_{0.09}$ alloy surface. (a)** Backscattered electron SEM micrograph at 18 kV, 3.2 nA and **(b)** compositional line profile across the secondary phase in at%. Six different EDS line profiles were taken from different second phase spots and their adjacent host alloy matrix, and the compositional information is presented as an average of 625 data points shown in **(c).**

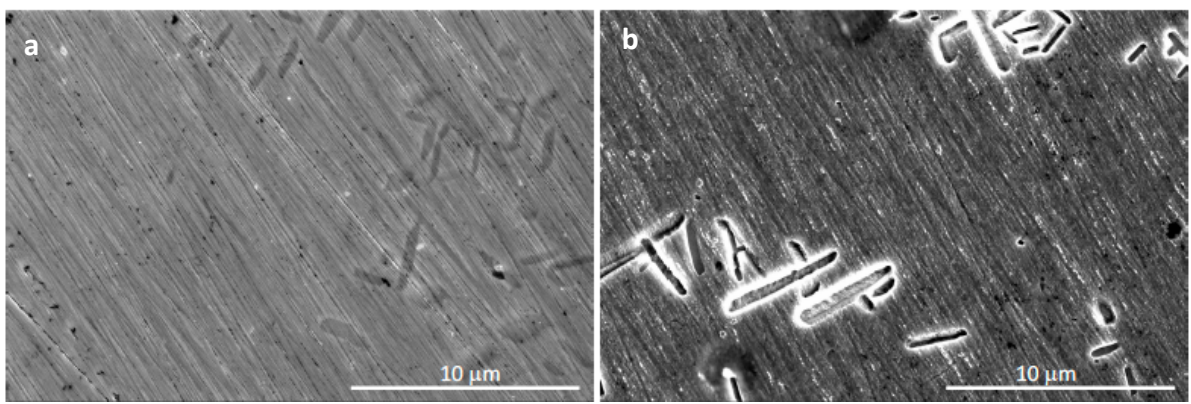

**Figure S9. SEM micrographs of (FeCoNi)$_{0.81}$Cr$_{0.10}$Al$_{0.09}$ alloy surface before (a) and after (b) chronoamperometry at 150 mV.** The precipitate particles are the 2$^{nd}$ phase $\beta$-NiAl, which have undergone dissolution during electrochemical exposure.

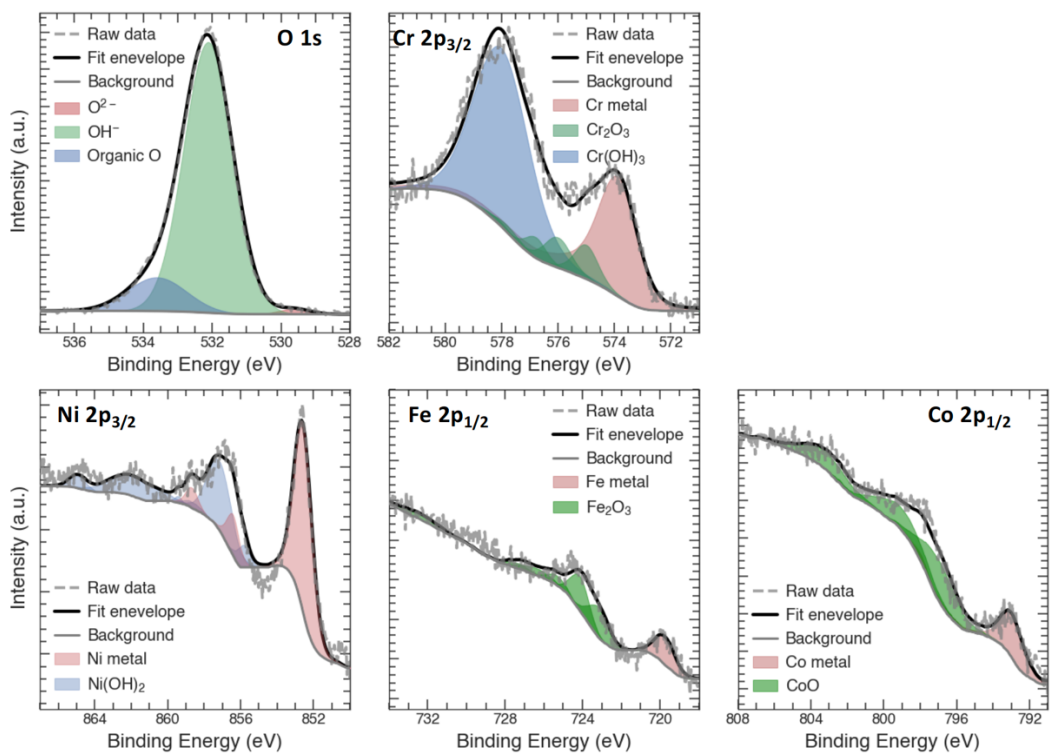

**Figure S10. XPS peak fitting and deconvolutions for (FeCoNi)$_{0.90}$Cr$_{0.10}$.**

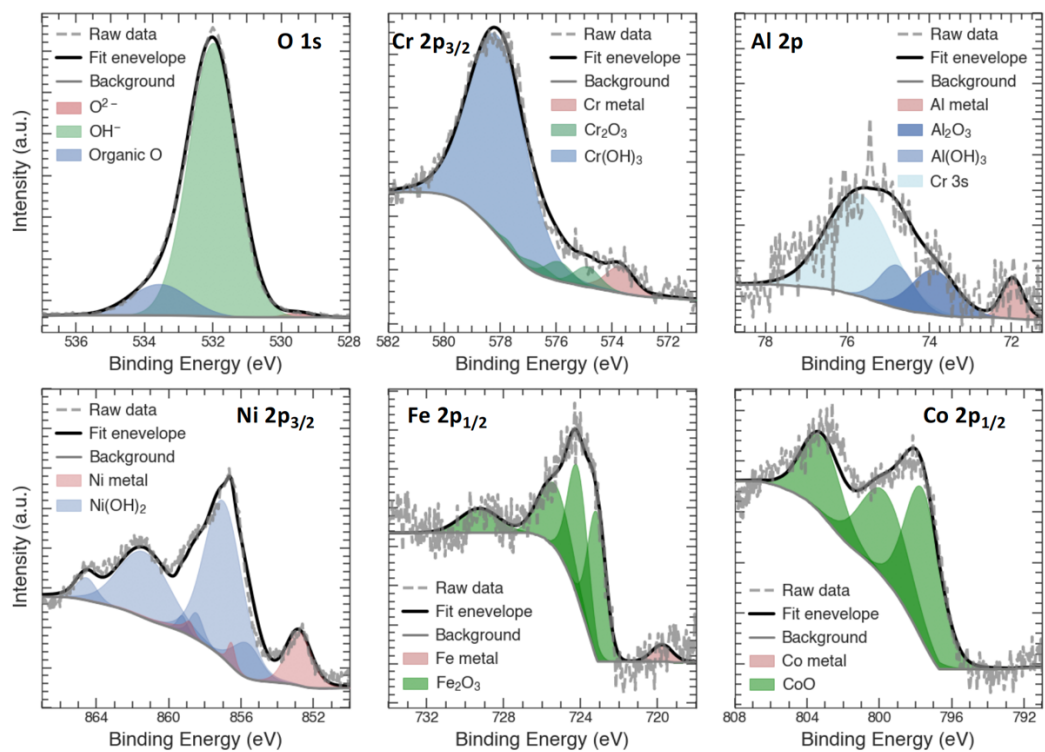

**Figure S11. XPS peak fitting and deconvolutions for (FeCoNi)$_{0.87}$Cr$_{0.10}$Al$_{0.03}$.**

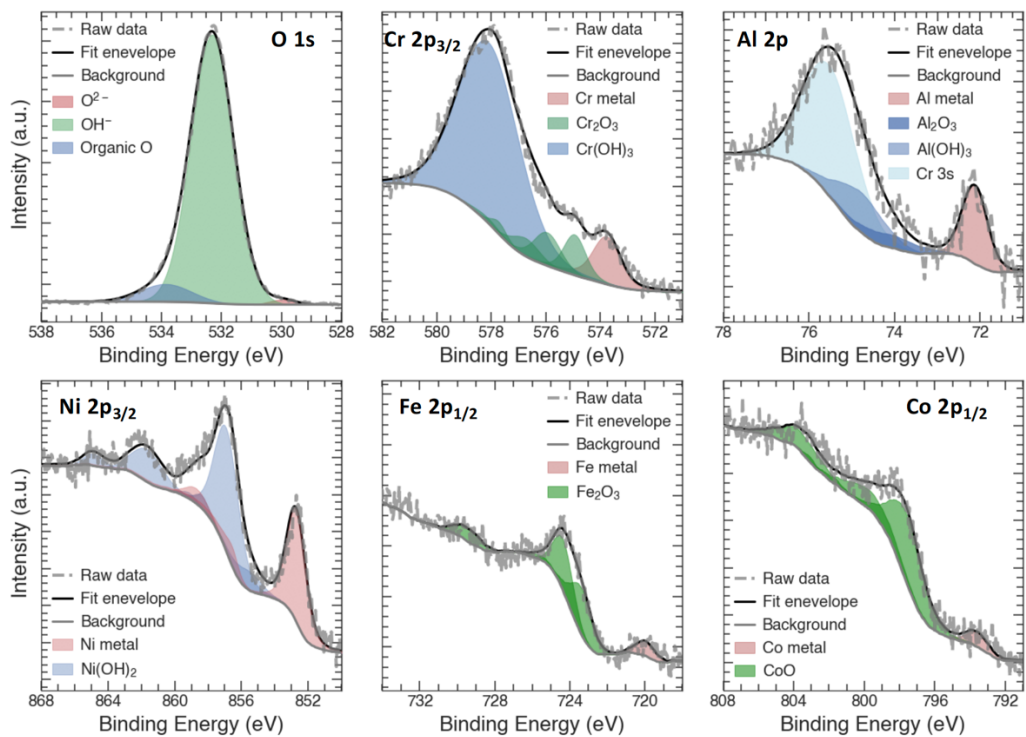

**Figure S12. XPS peak fitting and deconvolutions for (FeCoNi)$_{0.84}$Cr$_{0.10}$Al$_{0.06}$.**

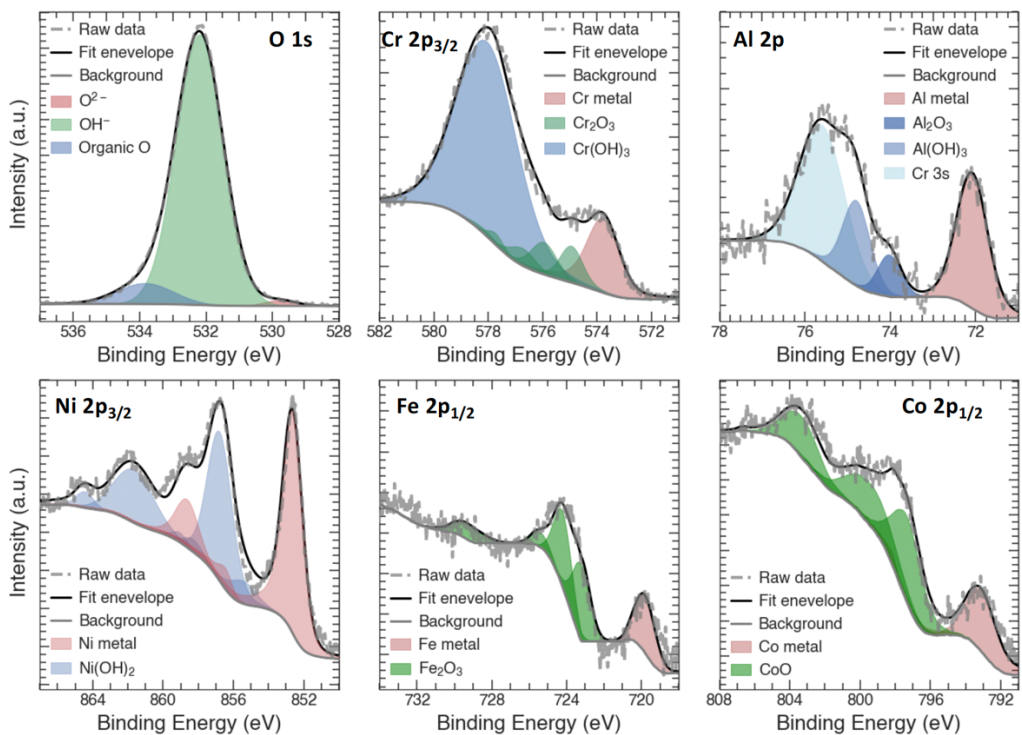

**Figure S13. XPS peak fitting and deconvolutions for (FeCoNi)$_{0.81}$Cr$_{0.10}$Al$_{0.09}$.**

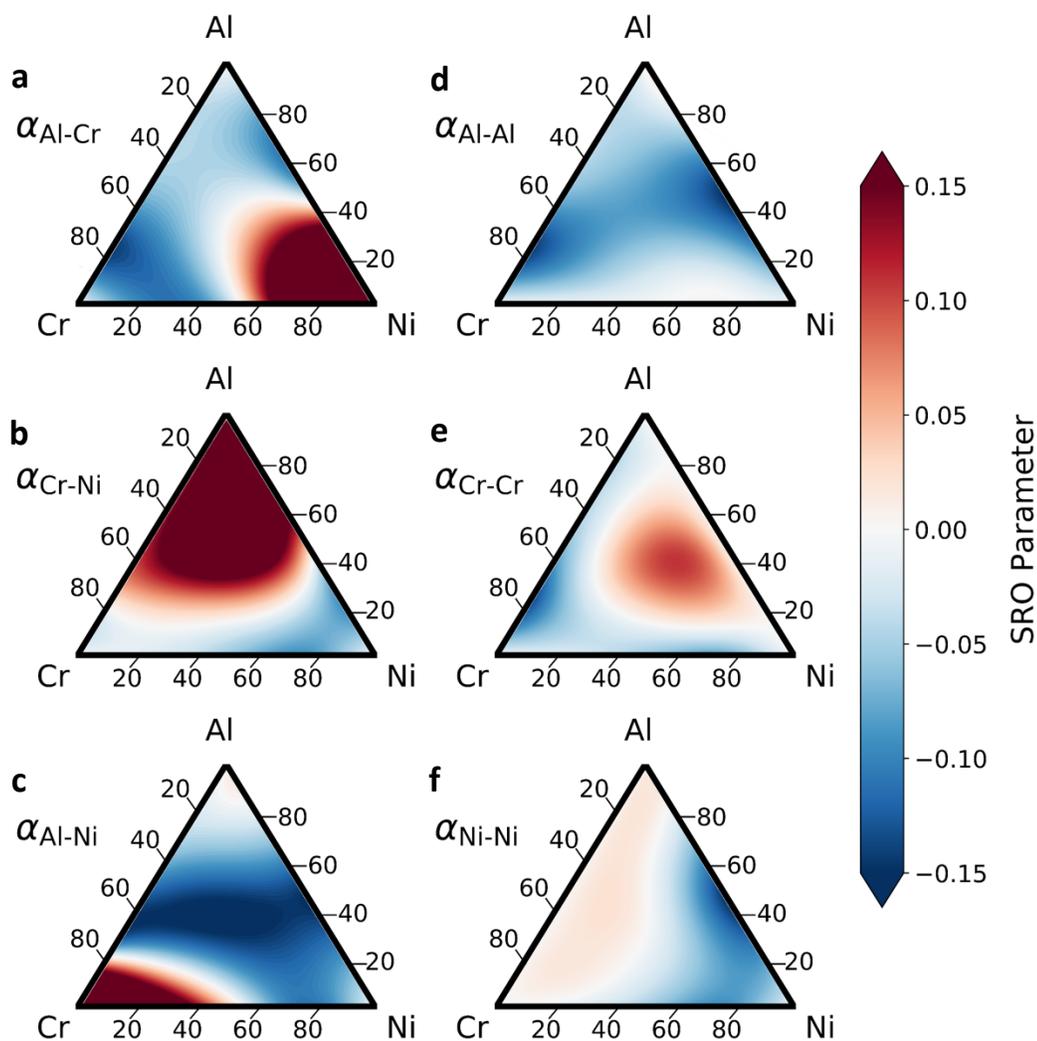

**Figure S14. Additional first principles cluster expansion and Monte Carlo results for ternary Ni₁₋ₓ₋yCrₓAl_y alloys.** Ternary diagrams of mixed-species **(a-c)** and same-species **(d-f)** W-C SRO parameter in the Al-Cr-Ni alloy system across the full phase space. Each binary subsystem of this ternary shows ordered compounds in its ground state phase diagram. This is reflected in the ordering-type SRO present along each binary edge for its respective W-C SRO pair.

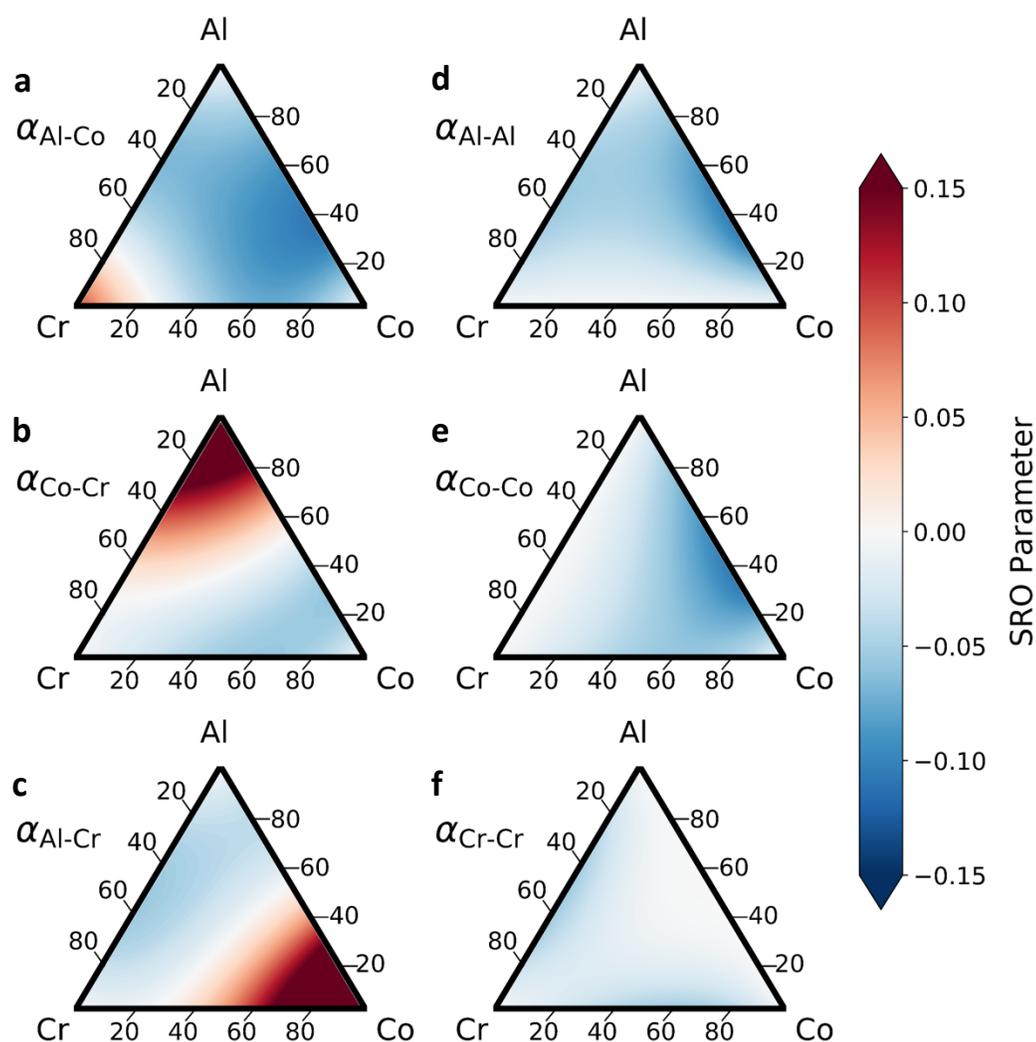

**Figure S15. Additional first principles cluster expansion and Monte Carlo results for ternary Co$_{1-x-y}$Cr$_x$Al$_y$ alloys.** Ternary diagrams of mixed-species **(a-c)** and same-species **(d-f)** W-C SRO parameter in the Al-Cr-Ni alloy system across the full phase space. Each binary subsystem of this ternary shows ordered compounds in its ground state phase diagram. This is reflected in the ordering-type SRO present along each binary edge for its respective W-C SRO pair.

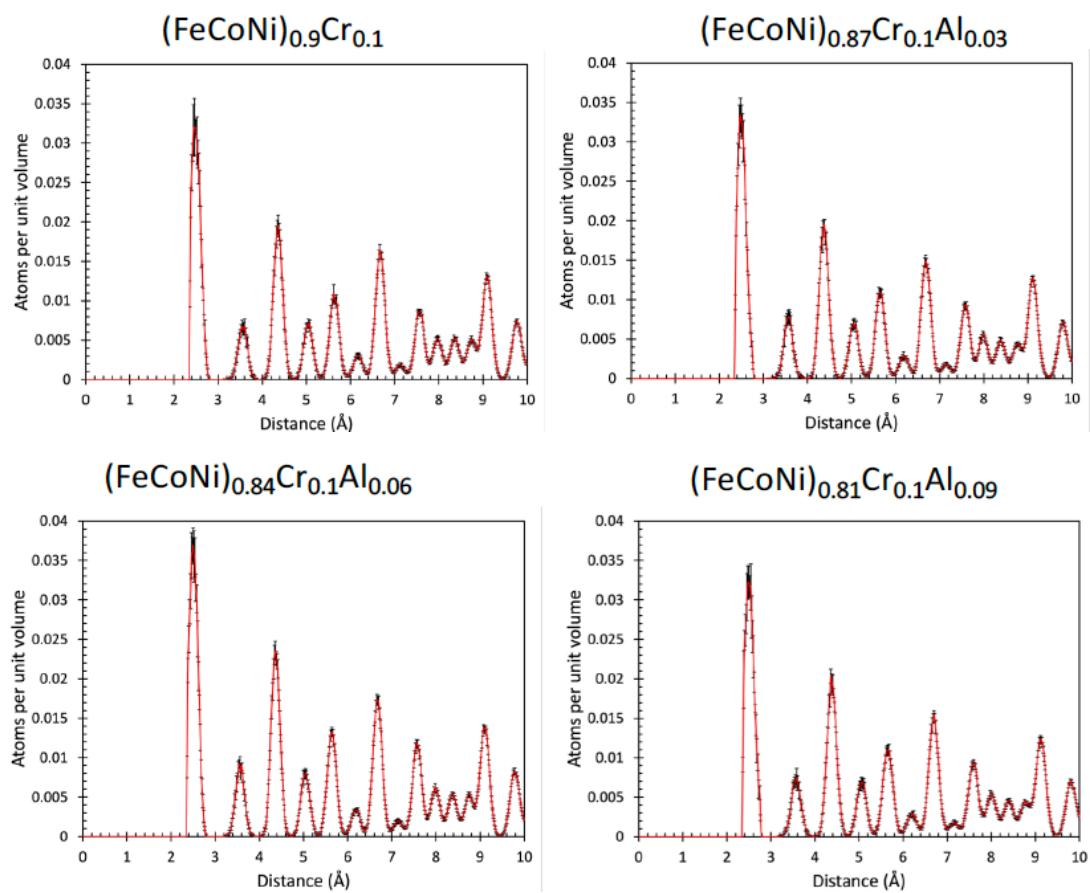

**Figure S16. Average density of Cr-Cr pairs in the 20 x 20 x 20, 32,000 atom Reverse Monte Carlo simulations of (FeCoNi)$_{1-y}$Cr$_{0.10}$Al$_y$.** Each data set is an average of 10 simulations.

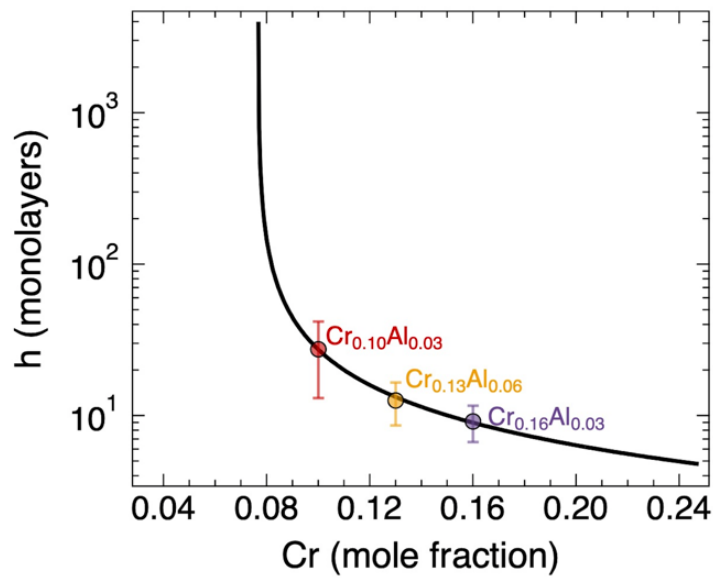

**Figure S17. Percolation theory fit to the *h*-value data shown in Fig. 1C (chronoamperometry at 150 mV) for the (FeCoNi)$_{0.87}$Cr$_{0.10}$Al$_{0.03}$ , (FeCoNi)$_{0.81}$Cr$_{0.13}$Al$_{0.06}$ and (FeCoNi)$_{0.81}$Cr$_{0.16}$Al$_{0.03}$ alloys.** The equation of the curve is $h = 1.015(p_c(h) - 0.077)^{-0.878}$.

**Table S1. Cationic fractions calculated from XPS peak fits and normalizations using the relative sensitivity factors provided in Table S2.**

| Alloy | Fe cations | Co cations | Ni cations | Cr cations | Al cations |
|---|---|---|---|---|---|
| $[FeCoNi]_{0.90}Cr_{0.10}$ | 0.20 | 0.32 | 0.24 | 0.24 | - |
| $[FeCoNi]_{0.87}Cr_{0.10}Al_{0.03}$ | 0.06 | 0.25 | 0.38 | 0.28 | 0.03 |
| $[FeCoNi]_{0.84}Cr_{0.10}Al_{0.06}$ | 0.12 | 0.23 | 0.27 | 0.33 | 0.04 |
| $[FeCoNi]_{0.81}Cr_{0.10}Al_{0.09}$ | 0.13 | 0.20 | 0.30 | 0.30 | 0.06 |
| $[FeCoNi]_{0.81}Cr_{0.13}Al_{0.06}$ | 0.10 | 0.23 | 0.30 | 0.32 | 0.05 |
| $[FeCoNi]_{0.84}Cr_{0.16}$ | 0.08 | 0.16 | 0.38 | 0.38 | - |
| $[FeCoNi]_{0.81}Cr_{0.16}Al_{0.03}$ | 0.10 | 0.22 | 0.26 | 0.39 | 0.02 |

**Table S2. Relative sensitivity factors provided by PHI for the XPS instrument Versaprobe III.**

| XPS Core Shell | Relative sensitivity factor |
|---|---|
| Fe $2p_{1/2}$ | 0.972 |
| Co $2p_{1/2}$ | 1.056 |
| Ni $2p_{3/2}$ | 2.309 |
| Cr $2p_{3/2}$ | 1.623 |
| Al $2p$ | 0.256 |
| Cr $3s$ | 0.080 |

**Table S3. XPS peak fit parameters of the core shells of Cr 2p$_{3/2}$, Fe 2p$_{1/2}$, Co 2p$_{1/2}$, Ni 2p$_{3/2}$, Al 2p, and O 1s comprising the passive film of [FeCoNi]$_{0.81}$Cr$_{0.10}$Al$_{0.09}$ after 10 ks of potentiostatic hold at +150 mV vs. SHE. A bar under the FWHM columns indicates that these species were not detected by our analysis.**

| Species | Ox. state | Multiplet 1 | | Multiplet 2 | | Multiplet 3 | | Multiplet 4 | | Multiplet 5 | | Multiplet 6 | |
|---|---|---|---|---|---|---|---|---|---|---|---|---|---|
| | | BE(eV) | FWHM | BE(eV) | FWHM | BE(eV) | FWHM | BE(eV) | FWHM | BE(eV) | FWHM | BE(eV) | FWHM |
| **Cr 2p$_{3/2}$** | | | | | | | | | | | | | |
| Cr metal | 0 | 573.7 | 1.36 | | | | | | | | | | |
| Cr$_2$O$_3$ | 3 | 574.95 | 1.11 | 575.98 | 1.11 | 576.74 | 1.11 | 577.74 | 1.11 | 578.15 | 1.11 | | |
| Cr(OH)$_3$ | 3 | 578.01 | 2.61 | | | | | | | | | | |
| **Ni 2p$_{3/2}$** | | | | | | | | | | | | | |
| Ni metal | 0 | 852.62 | 1.32 | 856.67 | 0.50 | 858.92 | 0.42 | | | | | | |
| NiO | 2 | 853.719 | - | 855.429 | - | | | 860.869 | - | 866.349 | - | | |
| Ni(OH)$_2$ | 2 | 855.51 | 1.60 | 856.86 | 2.50 | 858.30 | 0.78 | 859.09 | 0.63 | 860.08 | 3.25 | 864.38 | 1.34 |
| NiOOH | 2 | 854.61 | - | 856.13 | - | 861.54 | - | 867.119 | - | | | | |
| **Fe 2p$_{1/2}$** | | | | | | | | | | | | | |
| Fe metal | 0 | 719.81 | 1.31 | | | | | | | | | | |
| FeO | 2 | 721.81 | - | 723.11 | - | 724.31 | - | 725.51 | - | 728.81 | - | | |
| Fe$_2$O$_3$ | 3 | 723.21 | 1.08 | 724.21 | 1.11 | 725.41 | 1.12 | 726.51 | 1.08 | 727.41 | 1.4 | 729.31 | 1.86 |
| FeCr$_2$O$_4$ | 2 | 722.21 | - | 723.41 | - | 724.31 | - | 725.51 | - | 726.91 | - | | |
| FeOOH | 3 | 723.50 | - | 724.50 | - | 725.40 | - | 726.50 | - | 727.61 | - | 734.01 | - |
| **Al 2p** | | | | | | | | | | | | | |
| Al metal | 0 | 72.08 | 0.83 | | | | | | | | | | |
| Al$_2$O$_3$ | 3 | 74.10 | 0.85 | | | | | | | | | | |
| Al(OH)$_3$ | 3 | 74.80 | 0.85 | | | | | | | | | | |
| Cr 3s | - | 75.59 | 1.26 | | | | | | | | | | |
| **Co 2p$_{1/2}$** | | | | | | | | | | | | | |
| Co metal | 0 | 793.12 | 2.01 | 796.12 | 3.26 | 798.12 | 0.51 | | | | | | |
| CoO | 2 | 797.3 | 1.98 | 799.42 | 3.22 | 802.82 | 1.54 | 803.32 | 2.36 | | | | |
| Co(OH)$_2$ | 2 | 795.23 | - | 796.43 | - | 796.43 | - | 804.13 | - | | | | |
| **O 1s** | | | | | | | | | | | | | |
| O$^{2-}$(oxide) | 2 | 529.96 | 1.00 | | | | | | | | | | |
| OH- (hydroxide) | 2 | 532.03 | 1.68 | | | | | | | | | | |
| Organic O | 2 | 533.86 | 2.00 | | | | | | | | | | |

**Table S4. CSRO parameters determined from time of flight neutron scattering data.**

$\alpha_{Cr-Cr}$

| Cr-Cr | $(FeCoNi)_{0.9}Cr_{0.1}$ | $(FeCoNi)_{0.87}Cr_{0.1}Al_{0.03}$ | $(FeCoNi)_{0.84}Cr_{0.1}Al_{0.06}$ | $(FeCoNi)_{0.81}Cr_{0.1}Al_{0.09}$ |
|---|---|---|---|---|
| 1$^{st}$ NN | $0.010 \pm 0.002$ | $0.016 \pm 0.003$ | $0.020 \pm 0.002$ | $0.018 \pm 0.002$ |
| 2$^{nd}$ NN | $-0.005 \pm 0.003$ | $0.011 \pm 0.004$ | $0.020 \pm 0.003$ | $0.011 \pm 0.002$ |
| 3$^{rd}$ NN | $0.001 \pm 0.002$ | $0.005 \pm 0.001$ | $0.015 \pm 0.002$ | $0.009 \pm 0.002$ |
| 4$^{th}$ NN | $-0.001 \pm 0.003$ | $0.002 \pm 0.002$ | $0.006 \pm 0.003$ | $0.005 \pm 0.003$ |

$\alpha_{Cr-Al}$

| Cr-Al | $(FeCoNi)_{0.9}Cr_{0.1}$ | $(FeCoNi)_{0.87}Cr_{0.1}Al_{0.03}$ | $(FeCoNi)_{0.84}Cr_{0.1}Al_{0.06}$ | $(FeCoNi)_{0.81}Cr_{0.1}Al_{0.09}$ |
|---|---|---|---|---|
| 1$^{st}$ NN | - | $-0.14 \pm 0.03$ | $-0.17 \pm 0.02$ | $-0.18 \pm 0.02$ |
| 2$^{nd}$ NN | - | $-0.11 \pm 0.03$ | $-0.17 \pm 0.03$ | $-0.11 \pm 0.03$ |
| 3$^{rd}$ NN | - | $-0.05 \pm 0.03$ | $-0.13 \pm 0.01$ | $-0.10 \pm 0.02$ |
| 4$^{th}$ NN | - | $-0.01 \pm 0.04$ | $-0.05 \pm 0.02$ | $-0.058 \pm 0.009$ |

$\alpha_{Al-Al}$

| Al-Al | $(FeCoNi)_{0.9}Cr_{0.1}$ | $(FeCoNi)_{0.87}Cr_{0.1}Al_{0.03}$ | $(FeCoNi)_{0.84}Cr_{0.1}Al_{0.06}$ | $(FeCoNi)_{0.81}Cr_{0.1}Al_{0.09}$ |
|---|---|---|---|---|
| 1$^{st}$ NN | - | $0.005 \pm 0.002$ | $0.011 \pm 0.002$ | $0.019 \pm 0.002$ |
| 2$^{nd}$ NN | - | $0.001 \pm 0.002$ | $0.012 \pm 0.002$ | $0.014 \pm 0.005$ |
| 3$^{rd}$ NN | - | $0.002 \pm 0.001$ | $0.007 \pm 0.002$ | $0.011 \pm 0.002$ |
| 4$^{th}$ NN | - | $0.001 \pm 0.002$ | $0.003 \pm 0.002$ | $0.005 \pm 0.003$ |

$P_{Cr-Cr}$ extracted from $\alpha_{Cr-Cr}(p) = \frac{P_{Cr-Cr} - \bar{c}_{Cr}}{1 - \bar{c}_{Cr}}$, generally with the form $\alpha_{ii}(p) = \frac{P_{ii}(p) - \bar{c}_i}{1 - \bar{c}_i}$

| Cr-Cr | $(FeCoNi)_{0.9}Cr_{0.1}$ | $(FeCoNi)_{0.87}Cr_{0.1}Al_{0.03}$ | $(FeCoNi)_{0.84}Cr_{0.1}Al_{0.06}$ | $(FeCoNi)_{0.81}Cr_{0.1}Al_{0.09}$ |
|---|---|---|---|---|
| 1$^{st}$ NN | $10.9 \pm 0.2\%$ | $11.4 \pm 0.2\%$ | $11.8 \pm 0.2\%$ | $11.7 \pm 0.2\%$ |
| 2$^{nd}$ NN | $9.6 \pm 0.3\%$ | $11.0 \pm 0.3\%$ | $11.8 \pm 0.3\%$ | $11.0 \pm 0.2\%$ |
| 3$^{rd}$ NN | $10.1 \pm 0.2\%$ | $10.5 \pm 0.1\%$ | $11.4 \pm 0.2\%$ | $10.8 \pm 0.2\%$ |
| 4$^{th}$ NN | $9.9 \pm 0.2\%$ | $10.2 \pm 0.2\%$ | $10.5 \pm 0.3\%$ | $10.5 \pm 0.3\%$ |

$NN_{Cr-Cr}$ (Cr atoms per 12 Nearest Neighbors)

| Cr-Cr | $(FeCoNi)_{0.9}Cr_{0.1}$ | $(FeCoNi)_{0.87}Cr_{0.1}Al_{0.03}$ | $(FeCoNi)_{0.84}Cr_{0.1}Al_{0.06}$ | $(FeCoNi)_{0.81}Cr_{0.1}Al_{0.09}$ |
|---|---|---|---|---|
| 1$^{st}$ NN | $1.31 \pm 0.02$ | $1.37 \pm 0.03$ | $1.42 \pm 0.02$ | $1.40 \pm 0.02$ |
| 2$^{nd}$ NN | $1.15 \pm 0.03$ | $1.32 \pm 0.04$ | $1.41 \pm 0.03$ | $1.31 \pm 0.03$ |
| 3$^{rd}$ NN | $1.21 \pm 0.02$ | $1.25 \pm 0.01$ | $1.37 \pm 0.02$ | $1.30 \pm 0.02$ |
| 4$^{th}$ NN | $1.19 \pm 0.03$ | $1.22 \pm 0.02$ | $1.27 \pm 0.04$ | $1.25 \pm 0.03$ |

$\alpha_{Co-Cr}$

| Co-Cr | $(FeCoNi)_{0.9}Cr_{0.1}$ | $(FeCoNi)_{0.87}Cr_{0.1}Al_{0.03}$ | $(FeCoNi)_{0.84}Cr_{0.1}Al_{0.06}$ | $(FeCoNi)_{0.81}Cr_{0.1}Al_{0.09}$ |
|---|---|---|---|---|
| 1$^{st}$ NN | $-0.110 \pm 0.005$ | $-0.17 \pm 0.01$ | $-0.221 \pm 0.009$ | $-0.21 \pm 0.01$ |
| 2$^{nd}$ NN | $0.05 \pm 0.01$ | $-0.14 \pm 0.01$ | $-0.24 \pm 0.01$ | $-0.15 \pm 0.01$ |
| 3$^{rd}$ NN | $-0.006 \pm 0.006$ | $-0.058 \pm 0.008$ | $-0.176 \pm 0.007$ | $-0.116 \pm 0.006$ |
| 4$^{th}$ NN | $0.02 \pm 0.01$ | $-0.02 \pm 0.01$ | $-0.073 \pm 0.004$ | $-0.072 \pm 0.006$ |

$\alpha_{Co-Al}$

| Co-Al | $(FeCoNi)_{0.9}Cr_{0.1}$ | $(FeCoNi)_{0.87}Cr_{0.1}Al_{0.03}$ | $(FeCoNi)_{0.84}Cr_{0.1}Al_{0.06}$ | $(FeCoNi)_{0.81}Cr_{0.1}Al_{0.09}$ |
|---|---|---|---|---|
| 1$^{st}$ NN | - | $-0.20 \pm 0.03$ | $-0.224 \pm 0.009$ | $-0.238 \pm 0.006$ |
| 2$^{nd}$ NN | - | $-0.15 \pm 0.02$ | $-0.24 \pm 0.01$ | $-0.16 \pm 0.01$ |
| 3$^{rd}$ NN | - | $-0.06 \pm 0.01$ | $-0.18 \pm 0.01$ | $-0.134 \pm 0.004$ |
| 4$^{th}$ NN | - | $-0.02 \pm 0.02$ | $-0.08 \pm 0.01$ | $-0.08 \pm 0.01$ |

$\alpha_{Fe-Co}$

| Fe-Co | $(FeCoNi)_{0.9}Cr_{0.1}$ | $(FeCoNi)_{0.87}Cr_{0.1}Al_{0.03}$ | $(FeCoNi)_{0.84}Cr_{0.1}Al_{0.06}$ | $(FeCoNi)_{0.81}Cr_{0.1}Al_{0.09}$ |
|---|---|---|---|---|
| $1^{st}$ NN | $0.078 \pm 0.006$ | $0.143 \pm 0.009$ | $0.196 \pm 0.004$ | $0.197 \pm 0.007$ |
| $2^{nd}$ NN | $-0.034 \pm 0.004$ | $0.110 \pm 0.007$ | $0.208 \pm 0.007$ | $0.132 \pm 0.007$ |
| $3^{rd}$ NN | $0.009 \pm 0.002$ | $0.048 \pm 0.006$ | $0.156 \pm 0.005$ | $0.110 \pm 0.007$ |
| $4^{th}$ NN | $-0.010 \pm 0.005$ | $0.020 \pm 0.003$ | $0.062 \pm 0.006$ | $0.066 \pm 0.007$ |

$\alpha_{Ni-Co}$

| Ni-Co | $(FeCoNi)_{0.9}Cr_{0.1}$ | $(FeCoNi)_{0.87}Cr_{0.1}Al_{0.03}$ | $(FeCoNi)_{0.84}Cr_{0.1}Al_{0.06}$ | $(FeCoNi)_{0.81}Cr_{0.1}Al_{0.09}$ |
|---|---|---|---|---|
| $1^{st}$ NN | $0.108 \pm 0.004$ | $0.18 \pm 0.01$ | $0.262 \pm 0.005$ | $0.265 \pm 0.006$ |
| $2^{nd}$ NN | $-0.052 \pm 0.008$ | $0.141 \pm 0.009$ | $0.279 \pm 0.006$ | $0.181 \pm 0.005$ |
| $3^{rd}$ NN | $0.008 \pm 0.003$ | $0.061 \pm 0.004$ | $0.207 \pm 0.006$ | $0.148 \pm 0.004$ |
| $4^{th}$ NN | $-0.012 \pm 0.005$ | $0.025 \pm 0.004$ | $0.077 \pm 0.007$ | $0.088 \pm 0.005$ |

$\alpha_{Ni-Fe}$

| Ni-Fe | $(FeCoNi)_{0.9}Cr_{0.1}$ | $(FeCoNi)_{0.87}Cr_{0.1}Al_{0.03}$ | $(FeCoNi)_{0.84}Cr_{0.1}Al_{0.06}$ | $(FeCoNi)_{0.81}Cr_{0.1}Al_{0.09}$ |
|---|---|---|---|---|
| $1^{st}$ NN | $-0.055 \pm 0.005$ | $-0.106 \pm 0.007$ | $-0.157 \pm 0.004$ | $-0.170 \pm 0.006$ |
| $2^{nd}$ NN | $0.028 \pm 0.005$ | $-0.085 \pm 0.004$ | $-0.168 \pm 0.007$ | $-0.117 \pm 0.004$ |
| $3^{rd}$ NN | $-0.005 \pm 0.003$ | $-0.035 \pm 0.002$ | $-0.124 \pm 0.004$ | $-0.098 \pm 0.003$ |
| $4^{th}$ NN | $0.007 \pm 0.004$ | $-0.014 \pm 0.004$ | $-0.048 \pm 0.004$ | $-0.059 \pm 0.006$ |

$\alpha_{Ni-Al}$

| Ni-Al | $(FeCoNi)_{0.9}Cr_{0.1}$ | $(FeCoNi)_{0.87}Cr_{0.1}Al_{0.03}$ | $(FeCoNi)_{0.84}Cr_{0.1}Al_{0.06}$ | $(FeCoNi)_{0.81}Cr_{0.1}Al_{0.09}$ |
|---|---|---|---|---|
| $1^{st}$ NN | - | $0.15 \pm 0.02$ | $0.193 \pm 0.009$ | $0.21 \pm 0.01$ |
| $2^{nd}$ NN | - | $0.10 \pm 0.02$ | $0.20 \pm 0.01$ | $0.15 \pm 0.01$ |
| $3^{rd}$ NN | - | $0.04 \pm 0.01$ | $0.149 \pm 0.009$ | $0.120 \pm 0.005$ |
| $4^{th}$ NN | - | $0.02 \pm 0.02$ | $0.06 \pm 0.02$ | $0.068 \pm 0.007$ |